\documentclass[pra,aps,reprint,superbib,superscriptaddress,twocolumn]{revtex4-1}
\usepackage{amsmath,bm,bbm}
\usepackage{amstext}
\usepackage{epsfig}
\usepackage{xcolor}
\usepackage{subfig}
\usepackage{graphicx}
\usepackage{multirow}
\usepackage{array}
\usepackage{tikz,pgfplots}
\usepackage{MnSymbol}
\usepackage{dsfont}
\usepackage{bbold}
\usepackage{physics}
\usepackage{float}
\usepackage{mathtools}
\usepackage{braket}
\usepackage[colorlinks=true,urlcolor=blue,citecolor=blue,linkcolor=blue]{hyperref}
\usepackage[toc]{appendix}
\definecolor{dgreen}{rgb}{0,.5,0}
\definecolor{dred}{rgb}{.7,.0,.0}

\newcommand{\be}{\begin{eqnarray}}
\newcommand{\ee}{\end{eqnarray}}

\DeclareMathOperator*{\argmin}{arg\,min}

\newcommand{\bX}{{\bm X}}
\newcommand{\bY}{{\bm Y}}

\newcommand{\bv}{{\bm v}}
\newcommand{\bV}{{\bm V}}
\newcommand{\bP}{{\bm P}}
\newcommand{\bI}{{\bm I}}
\newcommand{\bev}{{\bm e}}
\newcommand{\bg}{{\bm \gamma}}

\newcommand{\ie}{{\it{i.e.}}}
\newcommand{\calC}{{\mathcal{C}}}
\newcommand{\calE}{{\mathcal{E}}}
\newcommand{\cop}{\hat{c}}
\newcommand{\cdop}{\hat{c}^\dagger}
\newcommand{\dop}{\hat{d}}
\newcommand{\ddop}{\hat{d}^\dagger}
\newcommand{\Hc}{{\rm H.c.}}
\usepackage{xr}
\usepackage{appendix}

\begin{document}

\title{
Householder transformed density matrix functional embedding theory 
}
\author{Sajanthan Sekaran}
\affiliation{Laboratoire de Chimie Quantique,
Institut de Chimie, CNRS/Universit\'e de Strasbourg,
4 rue Blaise Pascal, 67000 Strasbourg, France}
\author{Masahisa Tsuchiizu}
\affiliation{
Department of Physics, Nara Women's University, Nara 630-8506, Japan}
\author{Matthieu Sauban\`{e}re}
\affiliation{
ICGM, Universit\'{e} de Montpellier, CNRS, ENSCM, Montpellier, France
}
\author{Emmanuel Fromager}
\affiliation{Laboratoire de Chimie Quantique,
Institut de Chimie, CNRS/Universit\'e de Strasbourg,
4 rue Blaise Pascal, 67000 Strasbourg, France}


\begin{abstract}

Quantum embedding based on the (one-electron reduced) density matrix is
revisited by means of the unitary Householder
transformation. While being exact and equivalent to (but
formally simpler than) density matrix
embedding theory (DMET) in the non-interacting case,
the resulting Householder transformed density matrix functional embedding
theory (Ht-DMFET) preserves, by construction, the single-particle
character of the bath when electron correlation is introduced. In
Ht-DMFET, the projected ``impurity+bath''
cluster's Hamiltonian (from which approximate local properties of the interacting lattice can be
extracted) becomes an explicit functional of the density matrix.    
In the spirit of single-impurity DMET, we consider in this work a closed (two-electron) cluster
constructed from the full-size non-interacting density matrix. When the
(Householder transformed) interaction on the bath site is taken into account, 
per-site energies obtained for the half-filled one-dimensional
Hubbard lattice match almost perfectly the exact Bethe
Ansatz results in all correlation regimes. 
In the strongly
correlated regime, the results deteriorate away from half-filling.
This can be related to the electron number fluctuations in the
(two-site) cluster which are not described neither in Ht-DMFET nor in
regular DMET. As
expected, the per-site energies dramatically improve when increasing the
number of embedded impurities. Formal connections with density/density
matrix functional theories have been briefly discussed and should be explored further. Work is currently
in progress in this direction.
\end{abstract}

\maketitle

\section{Introduction}

Quantum embedding has emerged over the last two decades
as a viable strategy for modelling
strong electron 
correlation in large molecules and extended systems~\cite{IJQC20_Adam-Michele_embedding_special_issue}. The purpose
of an embedding procedure is to replace the original full-size
problem, for which an accurate solution to the Schr\"{o}dinger equation
is out of reach, by one or several simpler problems that preserve only a fragment of the original
system. The fragment, which can be
a single atomic site (often referred to as {\it impurity}) in a lattice, is embedded into a
formal bath that is supposed to mimic the effects of the impurity's 
environment. The mathematical construction of the bath depends on the 
the choice of basic variable in the embedding procedure. Obviously, this
choice is
not unique, which explains the diversity of embedding schemes in the
literature~\cite{IJQC20_Adam-Michele_embedding_special_issue}. In the well-established {\it dynamical mean-field theory}
(DMFT)~\cite{georges1992hubbard,georges1996limitdimension,kotliar2004strongly,held2007electronic,zgid2011DMFTquantum},
the so-called local Green function, which is evaluated on the impurity,
is the quantity of interest. In this case, the non-interacting sites of
the Anderson model (on which the Green function is mapped) represent the
bath. In the simplified two-site formulation of DMFT~\cite{potthoff2001two}, the latter reduces to a single
site. Note that the fragmentation of a system, which is central
in embedding calculations, allows for the combination of different
electronic structure methods. We can refer, for example, to {\it density functional
theory} (DFT) or the Green-function-based $GW$ method on top of
DMFT~\cite{kotliar2006reviewDMFT,sun_extended_2002,biermann2003first,Haule_2ble_counting_DMFT-DFT_2015,boehnke2016strong,werner2016dynamical,nilsson2017multitier},
but also {\it self-energy embedding theory}
(SEET)~\cite{kananenka2015systematically,lan2015communication,zgid2017finite,lan_generalized_2017}
or the dynamical configuration interaction
method~\cite{dvorak_quantum_2019}. Various combinations of DFT with many-body wave functions have also been explored
for
model~\cite{fromager2015exact,senjean2018site,senjean2018multiple,requist2019model}
and {\it ab initio} quantum chemical 
Hamiltonians~\cite{toulouse2004long,JCP19_Ferte_srDFT_on_top_pair_dens,gagliardi2016multiconfiguration,CR18_Truhlar_Multiconf_DFTs,CR15_Wesolowski_FDE,mosquera2019domain},
with the purpose of improving DFT in the description of strong electron
correlation.
\\

In recent years, {\it density matrix
embedding theory}
(DMET)~\cite{knizia2012density,knizia2013density,sun2016quantum,wu2019projected,JCTC20_Chan_ab-initio_DMET} has attracted an increasing
attention~\cite{bulik2014density,tsuchimochi2015density,welborn2016bootstrap,senjean2019projected,PRB21_Booth_effective_dynamics_static_embedding}.
Its drastic simplification of the bath (when compared to DMFT) explains
this success. 
Moreover,
unlike DMFT, DMET is originally a frequency-independent (and therefore
formally simpler) theory. 
Later on it has been extended to the description of
dynamical properties~\cite{booth2015spectral} and non-equilibrium
dynamics~\cite{kretchmer2018real}.
As shown in 
Refs.~\onlinecite{ayral2017dynamical,lee2019rotationally,fertitta2018rigorous,JCP19_Booth_Ew-DMET_hydrogen_chain,PRB21_Booth_effective_dynamics_static_embedding},
formal connections can be established between the two theories. Since, in DMET, the number of bath sites equals (at most) the number of
impurity sites, the Schr\"{o}dinger equation can be
solved accurately (if not exactly) for the reduced-in-size
``impurity+bath'' system.        
Despite its name, DMET falls in many ways into the category of {\it wave
function}-based methods. It is clear in its exact formulation, where
the many-body bath states arise from the Schmidt
decomposition of the full-system many-body wave function~\cite{knizia2012density}. We note in
passing that embedding theories similar in spirit to DMET but based on the exact factorization
of the (electronic) many-body wave
function have been proposed very recently~\cite{PRL20_Lacombe_embedding_via_EF,requist2019fock}. In standard
implementations of DMET, the Schmidt decomposition is applied to a
single Slater determinant $\Phi$. In this case, the bath simply consists of effective sites
(or orbitals) that can be determined numerically from the overlap matrix
between the fragment and the occupied orbitals in
$\Phi$~\cite{wouters2016practical}. In other words, the quantum partitioning
in standard DMET is done at the single-particle level~\cite{muhlbach2018quantum}.
The (one-electron reduced) density matrix comes into play in the
optimization of $\Phi$, through mapping constraints. As pointed out in
previous works~\cite{tsuchimochi2015density,ayral2017dynamical,PRB21_Booth_effective_dynamics_static_embedding},
representability issues may arise in this context, thus leading to
difficulties in the numerical robustness of the method. Note that
relaxed constraints have been used, like in {\it density embedding theory}
(DET)~\cite{bulik2014density}, where only site occupations (\ie, diagonal elements
of the density matrix) are mapped. A connection with DFT can be established in 
this case~\cite{mordovina2019self}.\\ 

The performance of DMET can in principle be improved systematically by
incorporating correlation into the
bath~\cite{knizia2012density,fan2015cluster,tsuchimochi2015density,hermes2019multiconfigurational}. 
The latter is then described by {\it many-body} states. Nevertheless, the
numerical efficiency and formal beauty of standard (mean-field-based)
DMET calculations definitely
rely on the one-electron character of the bath. 
We may actually wonder
if DMET can be made formally exact and systematically improvable
while preserving a single-particle quantum partitioning
picture. The recently proposed {\it energy-weighted} DMET
(EwDMET)~\cite{fertitta2018rigorous,JCP19_Booth_Ew-DMET_hydrogen_chain,PRB21_Booth_effective_dynamics_static_embedding}
is an elegant way to incorporate quantum fluctuations through
a (truncated) description of the effective dynamics. We want to follow a
different path where, ideally, the fully {\it static} character of DMET
would be preserved.  
The present work is a
first step toward this goal. More precisely, we rewrite the embedding as
a functional of the density matrix, thus bypassing the Schmidt decomposition of the reference
(correlated or not) full-system wave function. This is achieved {\it via} a unitary Householder
transformation~\cite{householder_unitary_1958}.
 In this approach, that
we refer to as {\it Householder transformed density-matrix functional
embedding theory} (Ht-DMFET), the bath orbital
becomes a simple and
analytical 
functional of the (possibly correlated) density matrix, thus greatly simplifying its
construction, even in the commonly used non-interacting (or mean-field) case. 
In this work, we describe in detail the embedding of a {\it single} impurity in the
one-dimensional (1D) Hubbard model. Note that we also implemented a multiple-impurity
version of Ht-DMFET based on a
block Householder
transformation~\cite{AML99_Rotella_Block_Householder_transf}. Its
detailed description will be given in a separate paper.\\

The present paper is organized as follows. After introducing the Householder
transformation in Sec.~\ref{subsec:HHt}, we show how it can be applied to 
the density matrix in order to derive an alternative 
embedding procedure (Sec.~\ref{subsec:HHt_dens_mat}). A comparison with the Schmidt
decomposition is made in Sec.~\ref{subsec:compar_Schmidt_decomp}.
The introduction into the theory of a correlation potential and related embedding constraints is also discussed (in Sec.~\ref{subsec:embedding_constraints}). Exact and approximate formulations
of Ht-DMFET are detailed in non-interacting (Sec.~\ref{subsec:embedding_for_MF}) and interacting (Sec.~\ref{subsec:correlated_embedding}) cases, respectively.    
The present (single-shot) implementation is summarized in
Sec.~\ref{subsec:summary_connections}. Its connection to DMET is also discussed. Results obtained for the 1D Hubbard model are presented in
Sec.~\ref{sec:results}. Conclusions and perspectives are finally given
in Sec.~\ref{sec:conclusions}.

\section{Theory}

For simplicity, embedding strategies will be discussed in the following for the
1D
Hubbard Hamiltonian,
\be\label{eq:uniform_hamilt_1D}
\hat{H}=-t\sum_{\sigma=\uparrow,\downarrow}\sum^{L-1}_{ i= 0
}\left(\hat{c}^\dagger_{i\sigma}\hat{c}_{(i+1)\sigma}+
{\rm H.c.}
\right)
+U\sum^{L-1}_{i=0}\hat{n}_{i\uparrow}\hat{n}_{i\downarrow},
\ee
where $\hat{n}_{i\sigma}=\hat{c}_{i\sigma}^\dagger\hat{c}_{i\sigma}$,
$\hat{c}^\dagger_{L\sigma}\equiv\hat{c}^\dagger_{0\sigma}$
($-\hat{c}^\dagger_{0\sigma}$) when periodic (antiperiodic) boundary
conditions are used,
$t$ is the hopping parameter, and $U$ is the on-site repulsion parameter. A {\it
single} impurity site (labelled as $i=0$) will be embedded. Extensions to 
more general (higher-dimension or quantum chemical) Hamiltonians with multiple impurities are left for future work.

\subsection{Householder transformation}\label{subsec:HHt}

Let us consider two different column vectors $\bX$ and $\bY$ with 
the same norm
\be\label{eq:normX_eq_normY}
\abs{\bX}=\sqrt{\bX^\dagger\bX}=\sqrt{\bY^\dagger\bY}=\abs{\bY}.
\ee 
What we will refer to as Householder
transformation~\cite{householder_unitary_1958} in the following 
simply corresponds to the reflection that transforms $\bX$
into $\bY$. As shown in Appendix~\ref{appendix:YeqPX} and sketched in Fig.~\ref{fig:HH_reflection_fig}, its matrix
representation $\bP$, which fulfills 
\be\label{eq:YeqPX}
\bY=\bP\bX,
\ee
can be written explicitly as follows, 
\be\label{eq:P_def}
{\bP}=\bI-2{\bv}{\bv}^\dagger,
\ee
where $\bI$ is the identity matrix and
\be\label{eq:HH_vec}
\bv=\dfrac{\bX-\bY}{\abs{\bX-\bY}}
\ee
is the normalized {\it Householder vector}. Note that $\bP$ is
both unitary and hermitian.
Note that real algebra will be used throughout the paper. As shown in
Fig.~\ref{fig:HH_reflection_fig}, the Householder vector is, by
construction, orthogonal to the reflection's hyperplane.\\

Let us now connect the Householder transformation
to density
matrix embedding. Following the basic idea of
DMET~\cite{knizia2012density}, we select an atomic site (say
$i=0$), referred to as impurity, and reduce drastically the
size of its environment in the lattice, in order to describe electron correlation. For that
purpose, we use the density
matrix elements that connect the impurity to the other lattice sites. All these
elements will be collected into the above-mentioned column vector $\bX$,
\ie,
\be\label{eq:Xdagger}
\bX^\dagger=\left[\gamma_{00},\gamma_{10},\ldots,\gamma_{i0},\ldots\right],
\ee        
where, as we focus here on singlet ground states, the (spin) density matrix elements
are denoted without spin indices: 
\be\label{eq:1RDM_notation}
{\gamma}_{ij}
=\left\langle\hat{c}^\dagger_{i\uparrow}\hat{c}_{j\uparrow}\right\rangle=
\left\langle\hat{c}^\dagger_{i\downarrow}\hat{c}_{j\downarrow}\right\rangle.
\ee 
In the reduced-in-size system, which we refer to as {\it
Householder cluster}, we would like to preserve the
(spin up or down) ocupation $\gamma_{00}$ of the impurity site. The 
simplest cluster we can think of is a dimer consisting of the impurity
and a bath site that needs to be determined. Ideally, we would
like the cluster to be, within the density matrix, disconnected from
its environment (whether this can actually be achieved or not will be
discussed later on). On that basis, we define our
Householder transformation-based embedding as follows:    
\be\label{eq:Ydagger}
\bX^\dagger\overset{\bP}{\longrightarrow}
\bY^\dagger=\left[\gamma_{00},\xi,0,0,\ldots,0,\ldots\right],
\ee
where all but the first two rows of $\bY$ are set to zero and the
constraint of Eq.~(\ref{eq:normX_eq_normY}) reads
\be\label{eq:xi_square}
\xi^2=\sum_{j>0}\gamma^2_{j0}.
\ee
As a result,
\be
\begin{split}
\abs{\bX-\bY}^2&=\left(\gamma_{10}-\xi\right)^2+\sum_{j>1}\gamma^2_{j0}
\\
&=\left(\gamma_{10}-\xi\right)^2+\xi^2-\gamma^2_{10},
\\
&=2\xi\left(\xi-\gamma_{10}\right).
\end{split}
\ee
Obviously, $\xi\neq0$, otherwise the impurity would be disconnected from
the rest of the lattice (and there would be no need for an embedding). In the
special case where $\gamma_{j0}\overset{j>1}{=}0$, which means that the
impurity's nearest-neighboring site $i=1$ is the bath, the
transformation should still be defined, meaning that $\abs{\bX-\bY}$
should never vanish. The latter requirement fixes the sign of $\xi$,
thus leading to the final expressions [see Eqs.~(\ref{eq:P_def}),
(\ref{eq:HH_vec}), (\ref{eq:Xdagger}), and (\ref{eq:Ydagger})]:
\be\label{eq:Pij_exp}
P_{ij}=\delta_{ij}-2v_iv_j,
\ee
where
\be\label{eq:Householder_vec}
v_0&=&0,
\nonumber\\
v_1&=&\dfrac{\gamma_{10}-\xi}{\sqrt{2\xi\left(\xi-\gamma_{10}\right)}},
\nonumber\\
v_j&\overset{j\geq 2}{=}&\dfrac{\gamma_{j0}}{\sqrt{2\xi\left(\xi-\gamma_{10}\right)}},
\ee
and
\be\label{eq:xi_final_exp}
\xi=-{\rm sgn}\left({\gamma_{10}}\right)\sqrt{\sum_{j>0}\gamma^2_{j0}}.
\ee  

\begin{figure}
\begin{center}
\includegraphics[scale=0.5]{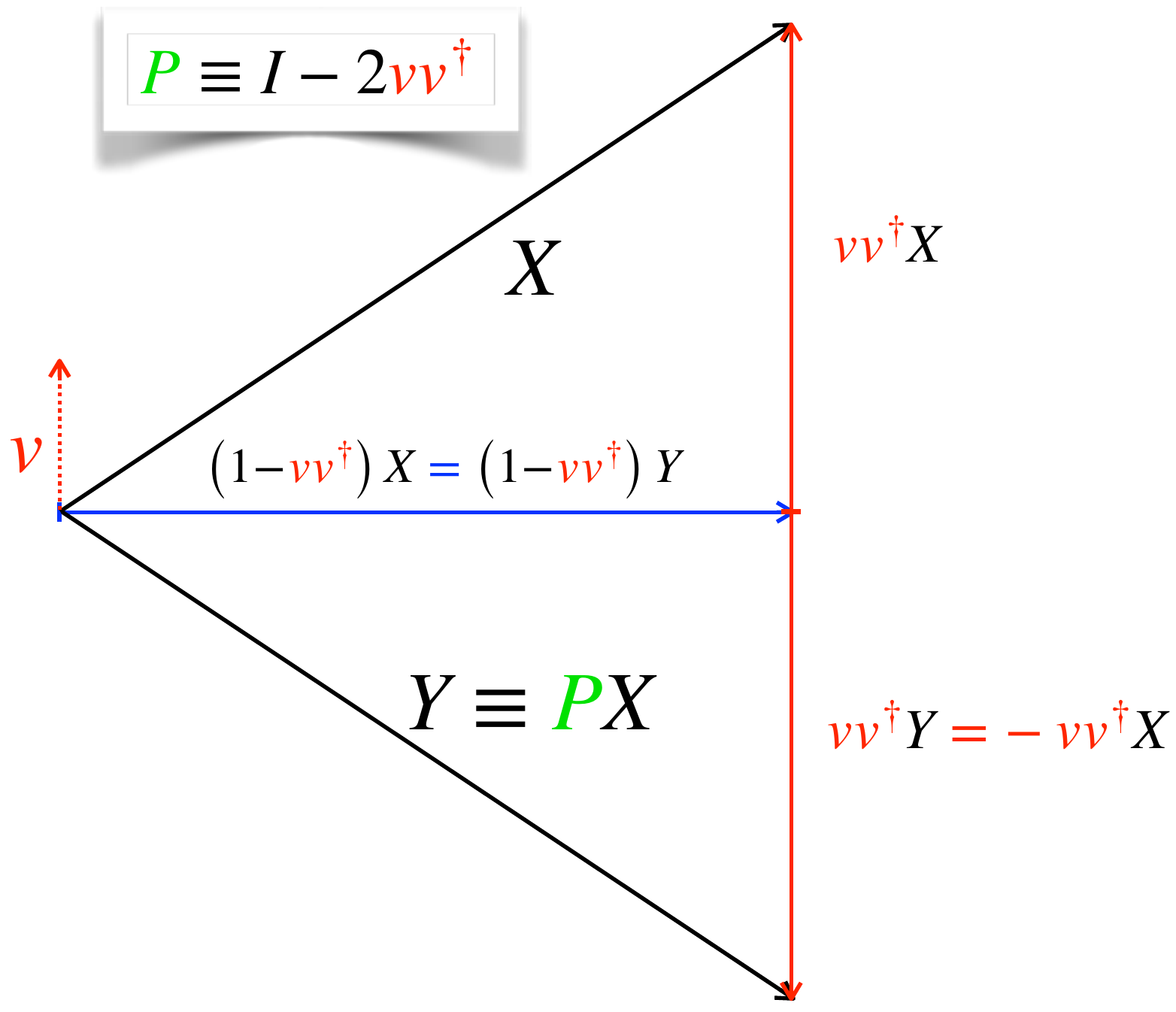}
\caption{Geometrical interpretation of the Householder transformation.
}
\label{fig:HH_reflection_fig}
\end{center}
\end{figure}

\begin{figure*}
\begin{center}
\includegraphics[scale=0.6]{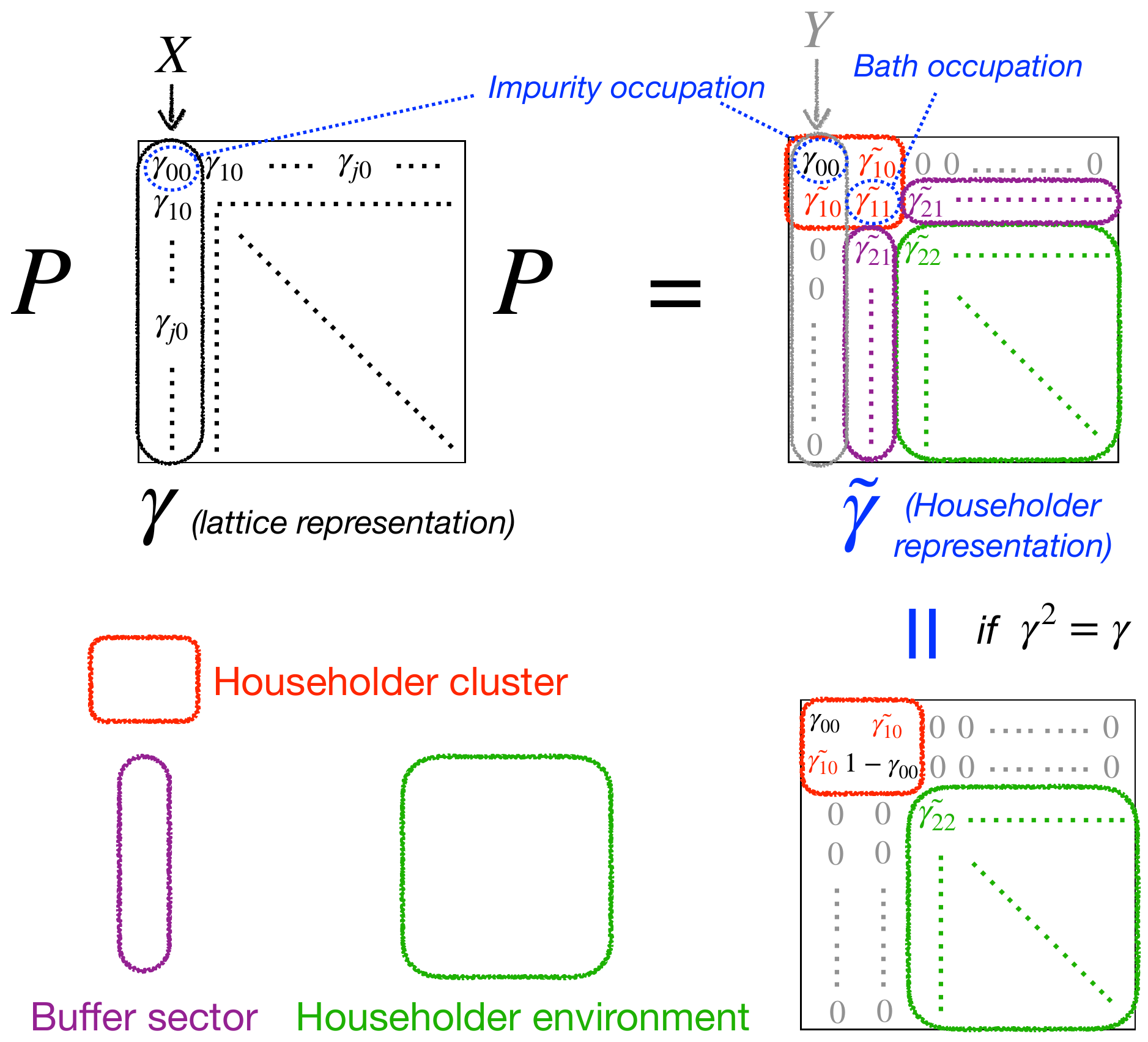}
\caption{Schematics of the Householder transformation applied to the
density matrix. 
}
\label{fig:schematics_HT_1RDM}
\end{center}
\end{figure*}
\subsection{Householder transformed density
matrix}\label{subsec:HHt_dens_mat}

\begin{figure}
\begin{center}
\includegraphics[scale=0.6]{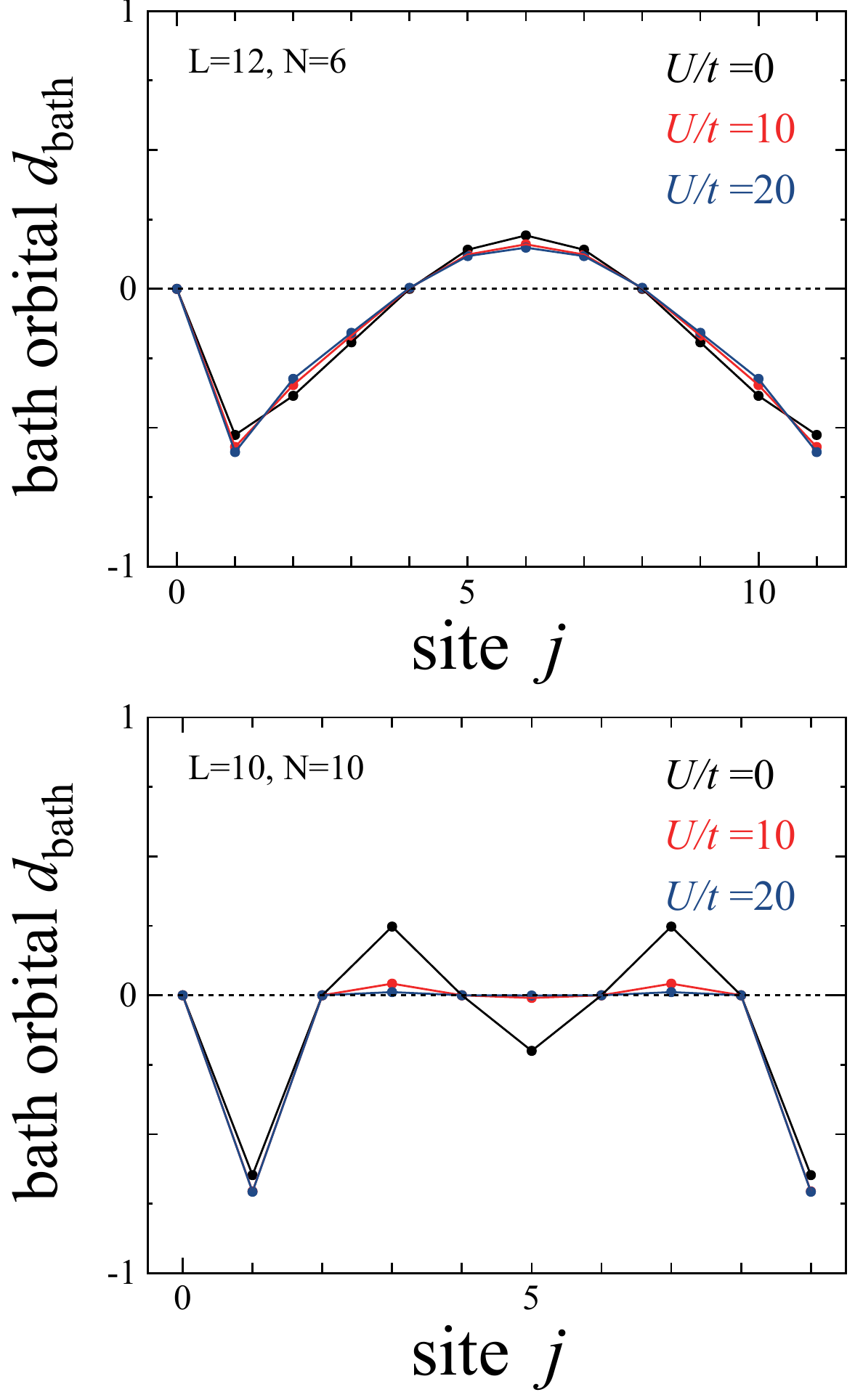}
\caption{Expansion of the exact (ground-state) bath orbital on the
lattice [see Eq.~(\ref{eq:bath_orbital})] for quarter-filled 12-site (top panel) and half-filled
10-site (bottom panel) Hubbard
rings in various correlation regimes. 
}
\label{fig:bath_orb}
\end{center}
\end{figure}

As sketched in Fig.~\ref{fig:schematics_HT_1RDM}, the Householder transformed density matrix can now be evaluated as
follows,
\be
\tilde{\bm \gamma}=\bP^\dagger {\bm \gamma}\bP=\bP{\bm \gamma}\bP,
\ee
or, equivalently,
\be
\tilde{\gamma}_{ij}=\expval{\hat{d}^\dagger_{i\sigma}\hat{d}_{j\sigma}},
\ee
where, according to Eqs.~(\ref{eq:Pij_exp}), (\ref{eq:Householder_vec}), and (\ref{eq:xi_final_exp}), the Householder transformed creation (annihilation) operators are
functionals of the density matrix elements $\gamma_{ij}$:
\be\label{eq:di_function_of_cj}
\begin{split}
\hat{d}^\dagger_{i\sigma}
&:=\sum_{j}P_{ij}\,\hat{c}^\dagger_{j\sigma}
\\
&=\hat{c}^\dagger_{i\sigma}-2v_i\sum_{j>0}v_j\,\hat{c}^\dagger_{j\sigma}.
\end{split}
\ee
We stress that the transformation leaves the impurity unchanged,
\be
\hat{d}^\dagger_{\rm imp}\equiv\hat{d}^\dagger_{0\sigma}=\hat{c}^\dagger_{0\sigma},
\ee
while the bath orbital is constructed explicitly from the environment
(in the lattice) of the impurity as follows,
\be\label{eq:bath_orbital}
\hat{d}^\dagger_{\rm
bath}\equiv\hat{d}^\dagger_{1\sigma}=\left(1-2v_1^2\right)\hat{c}^\dagger_{1\sigma}-2v_1\sum_{j>1}v_j\,\hat{c}^\dagger_{j\sigma}.
\ee
As shown in Fig.~\ref{fig:bath_orb}, for small Hubbard rings, 
the 
nearest neighbors of the impurity contribute the most to the bath.
Nevertheless, the delocalization of the latter over the lattice can be quite substantial,
in particular in the weakly correlated regime. The impact of correlation on
the bath varies with the lattice filling. For example, in the
quarter-filled 12-site ring, the deviation 
from the non-interacting bath orbital remains relatively small
when entering the strongly correlated regime (see the top panel of
Fig.~\ref{fig:bath_orb}). This is an important observation which makes
the use of a mean-field bath in conventional DMET
calculations~\cite{wouters2016practical} relevant. As
expected from Eq.~(\ref{eq:Householder_vec}) and
Appendix~\ref{appendix:Nc_half-filled_case} [see Eq.~(\ref{eq_appendix:1RDM_elt_zero_half-fill})], in the half-filled case,
only the lattice sites with odd indices contribute to the bath (see the
bottom panel of Fig.~\ref{fig:bath_orb}). Interestingly, in this case,
the bath delocalization reduces as correlation increases.\\         

Returning to the general theory, we note that the inverse
transformation (from the Householder representation to the lattice
one) simply reads
\be\label{eq:inverse_transf_less_explicit}
\sum_iP_{ki}\ddop_{i\sigma}=\sum_{ij}P_{ki}P_{ij}\cdop_{j\sigma}=\sum_{j}\delta_{kj}\cdop_{j\sigma}=\cdop_{k\sigma},
\ee
or, equivalently,
\be\label{eq:inverse_transf}
\cdop_{k\sigma}=\ddop_{k\sigma}-2v_k\sum_{i>0}v_i\ddop_{i\sigma}.
\ee
By construction, the first column (row) of the Householder transformed
density matrix equals zero outside the cluster [see
Fig.~\ref{fig:schematics_HT_1RDM}]. Indeed, according to Eqs.~(\ref{eq:Ydagger}) and (\ref{eq:Householder_vec}),
\be\label{eq:zero_first_column}
\begin{split}
\tilde{\gamma}_{j0}&=
\sum_{kl}P_{jk}\gamma_{kl}P_{l0}
\\
&=\sum_kP_{jk}\gamma_{k0}
\\
&=\left[\bP\bX\right]_j
\\
&=Y_j\overset{j\geq
2}{=}0.
\end{split}
\ee
Moreover, the occupation of the impurity is invariant under the
Householder transformation:
\be
\tilde{\gamma}_{00}=\sum_{kl}P_{0k}\gamma_{kl}P_{l0}={\gamma}_{00}.
\ee

\begin{figure}
\begin{center}
\includegraphics[scale=0.6]{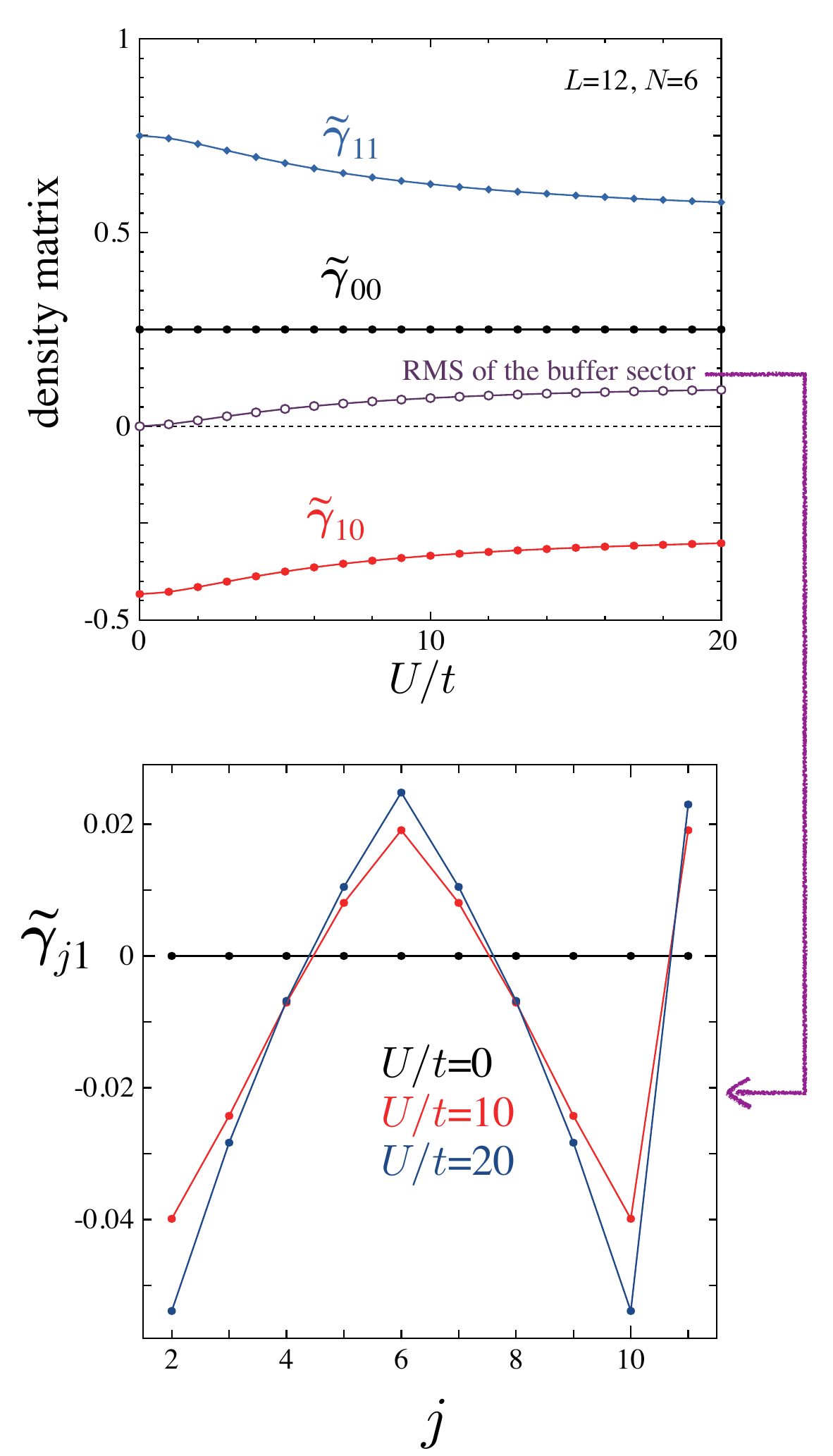}
\caption{ 
(Top) Exact (ground-state) Householder transformed density matrix
elements in the cluster sector and root mean square (RMS) of the
elements in the buffer sector, both plotted as functions of $U/t$ for a quarter-filled 12-site
Hubbard ring. (Bottom) Individual elements in the buffer sector for
various $U/t$ values. 
}
\label{fig:exact_results_quarter_filling}
\end{center}
\end{figure}

\begin{figure}
\begin{center}
\includegraphics[scale=0.6]{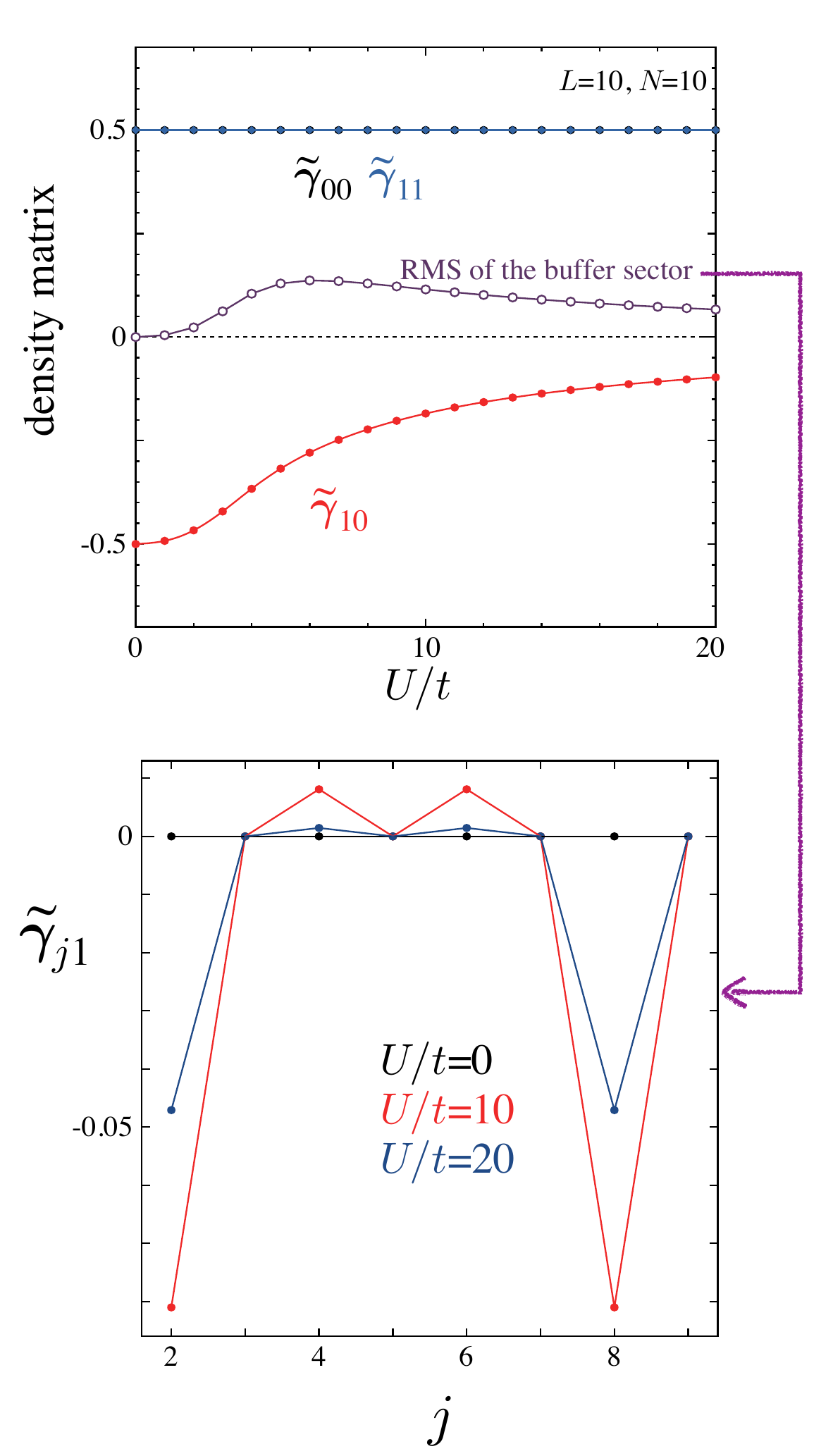}
\caption{Same as Fig.~\ref{fig:exact_results_quarter_filling} for a half-filled 10-site
Hubbard ring.
}
\label{fig:exact_results_half_filling}
\end{center}
\end{figure}

In the general (interacting) case, the
exact Householder cluster is {\it not} disconnected from its
environment. As shown in Figs.~\ref{fig:exact_results_quarter_filling} and \ref{fig:exact_results_half_filling} for small Hubbard rings, the Householder transformed (ground-state) density matrix has a nonzero
buffer sector $\{\tilde{\gamma}_{j1}\}_{j\geq 2}$, which is a signature 
of the cluster's entanglement with its environment. The existence of
such a buffer can be related to the deviation of the density matrix from
idempotency. Indeed, according to Eq.~(\ref{eq:zero_first_column}),   
\be\label{eq:calculating_gamma_square}
\begin{split}
\left[\tilde{\bm\gamma}^2\right]_{j0}=\sum_{k}\tilde{\gamma}_{jk}\tilde{\gamma}_{k0}&=\tilde{\gamma}_{j0}\tilde{\gamma}_{00}+\tilde{\gamma}_{j1}\tilde{\gamma}_{10}
\\
&\overset{j\geq
2}{=}\tilde{\gamma}_{j1}\tilde{\gamma}_{10},
\end{split}
\ee
thus leading to [see Eq.(\ref{eq:Ydagger})]
\be\label{eq:DM_buffer_region}
\tilde{\gamma}_{j1}\overset{j\geq
2}{=}\dfrac{\left[\tilde{\bm\gamma}^2\right]_{j0}}{\xi}
\ee
or, equivalently,
\be\label{eq:DM_buffer_region_idempotency_deviation}
\tilde{\gamma}_{j1}\overset{j\geq
2}{=}\dfrac{\left[\tilde{\bm\gamma}^2-\tilde{\bm\gamma}\right]_{j0}}{\xi}.
\ee
Another important observation [see Fig.~\ref{fig:exact_results_quarter_filling}] is that
the Householder cluster is, in general, an {\it open} subsystem. This
can also be related to the deviation from idempotency. 
Indeed, by considering the particular
case $j=1$ on the first line of Eq.~(\ref{eq:calculating_gamma_square}), it comes  
\be\label{eq:nbr_electrons_cluster}
\dfrac{\mathcal{N}^{\mathcal{C}}}{2}:=\tilde{\gamma}_{00}+\tilde{\gamma}_{11}=\dfrac{\left[\tilde{\bm\gamma}^2\right]_{10}}{\left[\tilde{\bm\gamma}\right]_{10}}=\dfrac{\left[\tilde{\bm\gamma}^2\right]_{10}}{\xi},
\ee
where $\mathcal{N}^{\mathcal{C}}$ is the total (spin-summed) number of
electrons in the cluster. Interestingly, when the interacting lattice is
half-filled, the fluctuations in the number of electrons within the
Householder cluster vanish and the latter contains exactly two electrons
for all $U/t$ values [see
Fig.~\ref{fig:exact_results_half_filling}]. Moreover, in the buffer
sector of the density matrix, elements with odd row (column) indices are
zero [see the bottom panel of Fig.~\ref{fig:exact_results_half_filling}].
Nevertheless, even in this
particular case, the cluster
remains connected to its Householder environment as long as the lattice is
interacting ($U/t\neq 0$). As proved analytically in
Appendix~\ref{appendix:Nc_half-filled_case}, these properties originate
from the hole-particle symmetry of the Hubbard Hamiltonian.

\subsection{Comparison with the Schmidt
decomposition}\label{subsec:compar_Schmidt_decomp}

Let us consider the determinant expansion of any (correlated or not) wave function $\Psi$ in
the Householder representation,
\be
\ket{\Psi}=\sum_{\mathcal{I}\mathcal{B}}\sum_{\mathcal{E}\in
\overline{{\mathcal{I}\mathcal{B}}}}
\lambda_{\mathcal{I}\mathcal{B}\mathcal{E}}
\ket{{\mathcal{I}\mathcal{B}\mathcal{E}}}
,
\ee
where $\ket{{\mathcal{I}\mathcal{B}\mathcal{E}}}=
\ket{\mathcal{I}}\ket{\mathcal{B}}\ket{\mathcal{E}}\equiv
\hat{\mathcal{I}}^\dagger\hat{\mathcal{B}}^\dagger\hat{\mathcal{E}}^\dagger\ket{\rm vac}$ consists of
impurity, bath, and Householder environment states, {\ie}, in a
second-quantized language, 
\be
\begin{split}
&\hat{\mathcal{I}}^\dagger=
\left(\hat{d}
_{0\uparrow}^\dagger
\right)
^{n^\mathcal{I}_{0\uparrow}}
\left(\dop
_{0\downarrow}^\dagger
\right)
^{n^\mathcal{I}_{0\downarrow}}
,
\\
&
\hat{\mathcal{B}}^\dagger
=
\left(\ddop
_{1\uparrow}
\right)
^{n^\mathcal{B}_{1\uparrow}}
\left(\ddop
_{1\downarrow}
\right)
^{n^\mathcal{B}_{1\downarrow}}
,
\\
&
\hat{\mathcal{E}}^\dagger
=
\left(\ddop
_{2\uparrow}
\right)
^{n^\mathcal{E}_{2\uparrow}}
\left(\ddop
_{2\downarrow}
\right)
^{n^\mathcal{E}_{2\downarrow}}
\ldots
\left(\ddop
_{(L-1)\uparrow}
\right)
^{n^\mathcal{E}_{(L-1)\uparrow}}
\left(\ddop
_{(L-1)\downarrow}
\right)
^{n^\mathcal{E}_{(L-1)\downarrow}}
,
\end{split}
\ee
$n^{\mathcal{I},\mathcal{B},\mathcal{E}}_{i\sigma}\in\left\{0,1\right\}$
being the occupation numbers.
The notation ``$\mathcal{E}\in
\overline{{\mathcal{I}\mathcal{B}}}$'' means that the summation over the
environment states is restricted to those preserving the total number of
electrons in the lattice. In order to evaluate the density matrix elements
$\tilde{\gamma}_{j0}=\bra{\Psi}\hat{d}^\dagger_{j\sigma}\hat{d}_{0\sigma}\ket{\Psi}$,
we need to look at the following overlaps: 
\be
\begin{split}
\mel{\mathcal{I}'\mathcal{B}'\mathcal{E}'}{\hat{d}^\dagger_{j\sigma}\hat{d}_{0\sigma}}{\mathcal{I}\mathcal{B}\mathcal{E}}
&\overset{j\geq 2}{\rightarrow} 
\delta_{\mathcal{B}'\mathcal{B}}
\mel{\mathcal{I}'}{\hat{d}_{0\sigma}}{\mathcal{I}}
\mel{\mathcal{E}'}{\hat{d}^\dagger_{j\sigma}}{\mathcal{E}}.
\end{split}
\ee
Those will contribute if the wave function exhibits, in its determinant expansion, an excitation from the
impurity to the environment. By imposing the constraint in Eq.~(\ref{eq:zero_first_column})
through the Householder transformation, we aim at removing such
excitations, thus forcing the impurity to exchange electrons only with
the bath spin-orbital $d_{1\sigma}$. 
The corresponding wave function can be expanded as follows, 
\be\label{eq:decomp_Householder_WF}
\ket{\Psi}=\sum_{\mathcal{B}}\ket{\mathcal{I}^\mathcal{B}\mathcal{B}}\sum_{\mathcal{E}^\mathcal{B}}\lambda_{\mathcal{I}^\mathcal{B}\mathcal{B}\mathcal{E}^\mathcal{B}}\ket{\mathcal{E}^\mathcal{B}},
\ee
where the occupation of the bath fixes the occupation of the impurity
(hence the notation $\mathcal{I}^\mathcal{B}$)
and the number of electrons in the environment. Eq.~(\ref{eq:decomp_Householder_WF}) can be seen as a {\it partial} Schmidt
decomposition~\cite{knizia2012density}. The latter would be complete if, for each bath state
$\mathcal{B}$, the environment were described by a single quantum state,
{\ie},
$\sum_{\mathcal{E}^\mathcal{B}}\lambda_{\mathcal{I}^\mathcal{B}\mathcal{B}\mathcal{E}^\mathcal{B}}\ket{\mathcal{E}^\mathcal{B}}\rightarrow
\lambda_\mathcal{B}\ket{\Psi_\mathcal{E}^\mathcal{B}}.
$
Note that, in the present formalism, the
bath refers to a one-electron quantum state, even for a correlated wave
function. In exact DMET,
the bath states are many-body quantum
states~\cite{knizia2012density}.\\  

As mentioned previously, in the general case, 
the buffer sector $\left\{\tilde{\gamma}_{j1}\right\}_{j\geq 2}$
is nonzero. Since 
\be
\begin{split}
\mel{\mathcal{I}'\mathcal{B}'\mathcal{E}'}{\hat{d}^\dagger_{j\sigma}\hat{d}_{1\sigma}}{\mathcal{I}\mathcal{B}\mathcal{E}}
&\overset{j\geq 2}{\rightarrow} 
\delta_{\mathcal{I}'\mathcal{I}}
\mel{\mathcal{B}'}{\hat{d}_{1\sigma}}{\mathcal{B}}
\mel{\mathcal{E}'}{\hat{d}^\dagger_{j\sigma}}{\mathcal{E}}
,
\end{split}
\ee
it means that the bath can in principle
exchange electrons with the Householder environment. In the particular
case where the lattice wave function 
$\Psi$ can be rewritten in a single-Slater-determinant form $\Phi$, which is
standard in practical DMET
calculations~\cite{wouters2016practical}, the density matrix becomes idempotent and the buffer sector
vanishes [see Eq.~(\ref{eq:DM_buffer_region_idempotency_deviation})].
Therefore, in this case, the cluster and its environment become
disentangled, as illustrated in Fig.~\ref{fig:schematics_HT_1RDM}. Moreover, the cluster contains exactly two electrons [see
Eq.~(\ref{eq:nbr_electrons_cluster})]. As a result, 
in the determinant expansion of Eq.~(\ref{eq:decomp_Householder_WF}),
each contributing cluster determinant $\ket{\mathcal{I}^\mathcal{B}\mathcal{B}}$ can be determined solely from the occupation of
the impurity (because the latter gives the occupation of the bath). By diagonalizing the (idempotent) Householder environment block of the density
matrix, we obtain a single Slater determinant
$\Phi_\mathcal{E}$ from which the full-system wave function can be determined as follows,
\be\label{eq:Schmidt_decomp_full_det}
\begin{split}
\ket{\Phi}&=\left(\sum_{\mathcal{I}}\lambda_{\mathcal{I}}\ket{\mathcal{I}\mathcal{B}^\mathcal{I}}\right)\ket{\Phi_\mathcal{E}}
\\
&=\sum_{\mathcal{I}}\lambda_{\mathcal{I}}\ket{\mathcal{I}}\ket{\mathcal{B}^\mathcal{I}\Phi_\mathcal{E}}
,
\end{split}
\ee
which is formally identical to a Schmidt decomposition where, in the
DMET terminology, the many-body bath
states are $\ket{\mathcal{B}^\mathcal{I}\Phi_\mathcal{E}}$ 
. The occupied orbitals in $\Phi_\mathcal{E}$ are usually referred to as {\it core}
embedding orbitals~\cite{wouters2016practical,kretchmer2018real}. Note that,
if we diagonalize the Householder (two-electron and idempotent) cluster block of the density matrix,
\be
\tilde{\bm\gamma}^{\mathcal{C}}=\begin{bmatrix}
\gamma_{00}&\xi\\
\xi & 1-\gamma_{00}
\end{bmatrix}
,
\ee
where the idempotency constraint reads
\be
\xi^2=\gamma_{00}\left(1-\gamma_{00}\right),
\ee 
we will obtain two spin-orbitals. The first one,
\be\label{eq:occ_in_cluster}
\ket{\mu_{\sigma}}=\dfrac{\xi}{\sqrt{1-\gamma_{00}}}\ket{c_{0\sigma}}+\sqrt{1-\gamma_{00}}\ket{d_{1\sigma}},
\ee
will be occupied, while its orthogonal counterpart (within the
cluster) will be unoccupied. In this particular
case, the cluster wave function can be written in a single-Slater-determinant
form,
\be\label{eq:Slater_det_cluster}
\sum_{\mathcal{I}}\lambda_{\mathcal{I}}\ket{\mathcal{I}\mathcal{B}^\mathcal{I}}\equiv\prod_{\sigma=\uparrow,\downarrow}\hat{\mu}^\dagger_\sigma\ket{\rm
vac}=\ket{\Phi^\calC},
\ee   
exactly like in standard (approximate) DMET [see
Eq.~(9) in Ref.~\onlinecite{wouters2016practical}]. In order to make the connection with DMET even more explicit, we can rewrite
Eq.~(\ref{eq:occ_in_cluster}) as follows,
\be
\begin{split}
\ket{\mu_{\sigma}}&=\braket{d_{0\sigma}|\mu_{\sigma}}\ket{d_{0\sigma}}+\braket{d_{1\sigma}|\mu_{\sigma}}\ket{d_{1\sigma}}
\\
&=\braket{c_{0\sigma}|\mu_{\sigma}}\ket{c_{0\sigma}}+\braket{d_{1\sigma}|\mu_{\sigma}}\ket{d_{1\sigma}},
\end{split}
\ee 
thus leading to
\be\label{eq:bath_DMET_way}
\begin{split}
\ket{d_{1\sigma}}&=\dfrac{1}{\sqrt{1-\braket{c_{0\sigma}|\mu_{\sigma}}^2}}\Big(\ket{\mu_{\sigma}}-\braket{c_{0\sigma}|\mu_{\sigma}}\ket{c_{0\sigma}}\Big)
\\
&=\dfrac{1}{\sqrt{1-\braket{c_{0\sigma}|\mu_{\sigma}}^2}}\sum_{j\geq
1}\braket{c_{j\sigma}|\mu_{\sigma}}\ket{c_{j\sigma}}.
\end{split}
\ee
If we now introduce, like in DMET, the overlap matrix
between all the occupied spin-orbitals $\nu_\sigma$ in $\Phi$
($\mu_\sigma$ being one of them),
\be
S_{\nu\nu'}=\braket{\nu_\sigma|c_{0\sigma}}\braket{c_{0\sigma}|\nu'_\sigma},
\ee    
it becomes clear that the Householder transformation is equivalent to
(but simpler than) the Schmidt
decomposition in the particular case of single
determinant wave functions. Indeed, by construction,
$\mu_\sigma$ is the only occupied spin-orbital that actually overlaps
with the impurity, thus leading to the simplified expression
\be
S_{\nu\nu'}=V_{\nu}\lambda^2V_{\nu'},
\ee   
where $V_{\nu}=\delta_{\nu\mu}$ and
$\lambda=\braket{c_{0\sigma}|\mu_{\sigma}}$. Thus, we recover from
Eq.~(\ref{eq:bath_DMET_way}) the
DMET expression for the bath [see Eq.~(11) in Ref.~\onlinecite{wouters2016practical}],
\be
\ket{d_{1\sigma}}=
\sum_{j\geq
1}
\sum^{\rm occ.}_\nu
\dfrac{\braket{c_{j\sigma}|\nu_{\sigma}}V_\nu}{\sqrt{1-\lambda^2}}\ket{c_{j\sigma}},
\ee 
where, according to Eq.~(\ref{eq:occ_in_cluster}),
\be
\lambda^2=\dfrac{\xi^2}{1-\gamma_{00}}=\gamma_{00}.
\ee
In summary, when the lattice is described with a single
Slater determinant, the Householder transformation 
and the Schmidt decomposition 
will lead to the same block-diagonalized density
matrix, each block corresponding to the Householder cluster and its
environment, respectively.
However, the Householder transformation is much simpler in this case
since the block-diagonalized structure is recovered automatically
({\it via} an analytical unitary transformation) from the density
matrix of the lattice.\\  

Obviously, the simplification in Eq.~(\ref{eq:Slater_det_cluster}) does
not hold anymore when electron correlation is introduced
{within the cluster}. Nevertheless, in this case, the (disentangled) wave function structure shown in
Eq.~(\ref{eq:Schmidt_decomp_full_det}) will be preserved and denoted later in Sec.~\ref{subsec:correlated_embedding}
as
follows:
\be
\ket{\Phi}=\ket{\Phi^{\mathcal{C}}}\ket{\Phi_{\mathcal{E}}}\rightarrow
\ket{\Psi^{\mathcal{C}}}\ket{\Phi_{\mathcal{E}}}. 
\ee
Note that, as shown in
Appendix~\ref{appendix:correlation_and_bath}, changes in the Householder cluster (and/or environment) sector of the
density matrix have no {\it direct} impact on the direction of the Householder
vector and therefore no impact on the bath. In other words, even though correlation within the
cluster and/or the environment modifies the density
matrix in the lattice representation, the correlated and original
(idempotent) density matrices will still share the
{\it same} Householder transformation [defined according to
Eqs.~(\ref{eq:Pij_exp}) and (\ref{eq:Householder_vec})] where the cluster
remains disconnected from its environment. 

\subsection{Embedding constraints, correlation potential, and connection
to DFT}\label{subsec:embedding_constraints}

In practical
(multiple-impurity~\cite{PRB21_Booth_effective_dynamics_static_embedding}) DMET
calculations, the bath is {\it indirectly} affected by the correlation
within the cluster through {\it ad hoc} density matrix mapping
constraints which can modify, {\it via} a (possibly nonlocal) correlation
potential operator acting on the full lattice~\cite{wouters2016practical,PRB20_Wu_DMET_local_corr_pot_fitting},
the
reference Slater determinant $\Phi$ of Eq.~(\ref{eq:Schmidt_decomp_full_det}). In the present formalism, it
would
induce a change in the Householder transformation and,
consequently, it would make the embedding procedure self-consistent.\\ 

As stressed in previous works~\cite{tsuchimochi2015density,ayral2017dynamical},
such a mapping cannot be exact, simply because, unlike a
correlated density matrix, the reference (non-interacting) one 
is idempotent.
From a formal point of view, rewriting the embedding procedure in the language of DFT, like in
DET~\cite{bulik2014density} or in the {\it self-consistent
density-functional embedding} (SDE) approach~\cite{mordovina2019self},
is more appealing because the (weaker) density-only constraint can be
fulfilled, in principle exactly. More precisely, if we combine Kohn--Sham
(KS) DFT~\cite{KStheory_1965} with a quasi-degenerate
perturbation
theory-like formalism~\cite{lindgren1986atomic,IJQC11_Christian-Brouder_QD-RSPT}, the
true interacting ground-state wave
function $\Psi_0$ can be described by its projection onto a
so-called {\it model}
Hilbert space $\mathcal{M}$. The latter would be constructed from
the (non-interacting KS) Householder cluster many-body states and the
core embedding KS orbitals. While the (ground-state) KS determinant
$\ket{\Phi_{\rm KS}}\equiv
\ket{\Phi_{\rm KS}^{\mathcal{C}}}\ket{\Phi_{\mathcal{E}}}$ belongs to $\mathcal{M}$
[see Eqs.~(\ref{eq:Schmidt_decomp_full_det}) and (\ref{eq:Slater_det_cluster})], it will of course not be the
case for the true solution $\Psi_0$. Its projection
$\ket{\Psi^{\mathcal{C}}}\ket{\Phi_{\mathcal{E}}}$ onto $\mathcal{M}$
can be determined, in principle exactly, {\it via} the diagonalization
of an effective 
Hamiltonian~\cite{lindgren1986atomic,IJQC11_Christian-Brouder_QD-RSPT}.
We can then restore $\Psi_0$ by applying a wave operator $\hat{\Omega}$, \ie,
\be
\ket{\Psi_0}\equiv \hat{\Omega}\left[\ket{\Psi^{\mathcal{C}}}\ket{\Phi_{\mathcal{E}}}\right]
,
\ee
thus making the embedding formally
exact. In practice, one may opt for a perturbation expansion of the
(local in this case)
correlation potential which can be determined order
by order from the density constraint $n_{\Psi_0}=n_{\Phi_{\rm KS}}$, by analogy with G\"{o}rling--Levy
perturbation
theory~\cite{gorling1994exact,IJQC95_Goerling-Levy_PT_density_constraints,JCP02_Ivanov_GL-PT_density_constraint}. Work is currently in progress in this direction.\\   

In the present work, where the embedding is applied to the uniform 1D
Hubbard lattice, the density profile is trivially obtained from the {\it
fixed} number
of electrons $N$ in the lattice. The local correlation potential is an arbitrary
constant and, therefore, what is referred to as {\it single-shot} embedding in the
DMET terminology~\cite{wouters2016practical} is sufficient. In other
words, as long as the number of electrons $N$ is fixed and we aim at
mapping only the correct filling $n=N/L$ onto the embedded impurity, there
is no correlation potential to optimize.    
\section{Single-shot density matrix functional
embedding}\label{sec:Ht-DMFET}

In the following, an exact formulation of Ht-DMFET is derived for a
non-interacting lattice (Sec.~\ref{subsec:embedding_for_MF}). On that basis, we construct an approximate
Ht-DMFET for the interacting lattice
(Sec.~\ref{subsec:correlated_embedding}). Finally, connections with
DMET and its extensions are made in
Sec.~\ref{subsec:summary_connections}.     
\subsection{Exact embedding in the non-interacting case
}\label{subsec:embedding_for_MF}

In the non-interacting (NI) case, which is equivalent
to the mean-field approximation in the uniform Hubbard model, the ground-state
per-site energy reads [see Eqs.~(\ref{eq:uniform_hamilt_1D}) and (\ref{eq:1RDM_notation})] 
\be
\label{eq:MF_ener}
e^{\rm NI}\left(n\right)= -4t\gamma_{10}
,
\ee
where we fix the number $N$ of electrons in the lattice and therefore the uniform filling
$n=N/L=2\gamma_{ii}$. 
Evaluating $\gamma_{10}$ requires solving the NI problem for the
full system, which is computationally affordable for a large number of
sites (we considered $L=400$ in this work). For that purpose, we
have to minimize the {\it total} NI energy
\be\label{eq:MF_total_ener_real-space}
E^{\rm NI}=
2\sum^{L-1}_{i,j=0}h_{ij}\gamma_{ij},
\ee
where 
\be\label{eq:hopping_matrix}
h_{ij}\overset{0<i<L-1}{=}-t\left[\delta_{j(i+1)}+\delta_{j(i-1)}\right],
\ee
and $h_{0(L-1)}\equiv \pm t$ (the sign depends on the
boundary conditions).
In practice, we simply need to diagonalize the hopping matrix ${\bm
h}\equiv \left\{h_{ij}\right\}$ and construct
the density matrix in the lattice representation from the occupied (orthonormal) eigen-spin-orbitals
$\ket{\kappa_{\sigma}}=\sum_{i}C_{i\kappa}\ket{c_{i\sigma}}$ as
follows,   
\be\label{eq:RDM_from_occupied_orbs}
\gamma_{ij}=\sum^{\rm occ.}_{\kappa}C_{i\kappa}C_{j\kappa}.
\ee
We propose in this section to reformulate the NI problem into an
embedded one. As such, it is in principle useless. However, when it
comes to introduce electron correlation, which is of course our ultimate
goal, this reformulation will provide a starting
point for the embedding. It will also suggest how the latter can be
systematically improved, as discussed further in Sec.~\ref{subsec:correlated_embedding}.\\  

Let us rewritte Eq.~(\ref{eq:MF_total_ener_real-space}) in the
Householder representation,
\be
E^{\rm NI}= 2\sum_{kl}\tilde{h}_{kl}\tilde{\gamma}_{kl},
\ee
where
\be\label{eq:tilde_f_ij}
\begin{split}
\tilde{h}_{kl}
&=\sum_{ij}P_{ki}h_{ij}P_{jl}
\\
&=h_{kl}
-2\sum_iv_i\left(v_kh_{il}+v_lh_{ik}\right)
+4v_kv_l\sum_{ij}v_ih_{ij}v_j
.
\end{split}
\ee
Since, in this representation, the (idempotent) density matrix can be split into
cluster and environment parts (see Fig.~\ref{fig:schematics_HT_1RDM}),
the same applies to the NI energy:  
\be
E^{\rm NI}
=E^{\rm NI}_{\calC}+E^{\rm NI}_{\calE}.
\ee
The cluster energy reads
\be
E^{\rm NI}_{\calC}=2\sum^1_{k,l=0}\tilde{h}_{kl}\tilde{\gamma}_{kl},
\ee
or, equivalently, 
\be
E^{\rm NI}_{\calC}=\mel{\Phi^\calC}{\hat{h}^{\calC}}{\Phi^\calC},
\ee
where $\Phi^\calC$ is the (single determinant) two-electron cluster wave function
introduced in Eq.~(\ref{eq:Slater_det_cluster}). 
The non-interacting Householder cluster and the Hubbard
dimer~\cite{carrascal2015hubbard,senjean2017local} 
have formally identical Hamiltonians, 
\be\label{eq:one_electron_hc}
\begin{split}
\hat{h}^{\calC}&=\sum_\sigma\sum^1_{i,j=0}\tilde{h}_{ij}\ddop_{i\sigma}\dop_{j\sigma}
\\
&\equiv
\sum_\sigma\left[
-\tilde{t}
\left(\ddop_{0\sigma}\dop_{1\sigma}+\Hc\right)+\tilde{\varepsilon}_1\ddop_{1\sigma}\dop_{1\sigma}\right],
\end{split}
\ee
where, according to Eq.~(\ref{eq:tilde_f_ij}),
\be\label{eq:tilde_t_exp}
\tilde{t}=-\tilde{h}_{01}=t
+2v_1\sum_iv_ih_{i0}
\ee
and
\be
\tilde{\varepsilon}_1=\tilde{h}_{11}=
-4v_1\sum_iv_i\left(h_{i1}-v_1\sum_jv_jh_{ij}\right).
\ee
On the other hand, the energy of the environment is an explicit functional of the
environment's density matrix
$\tilde{\gamma}^{\calE}\equiv\left\{\tilde{\gamma}_{ij}\right\}_{i>1,j>1}$: 
\be
E^{\rm NI}_{\calE}=2\sum^{L-1}_{k,l=2}\tilde{h}_{kl}\tilde{\gamma}_{kl}
\equiv 2\Tr\left[\tilde{\bm h}^\calE\tilde{\gamma}^{\calE}\right],
\ee
where $\Tr$ denotes the trace.
The total ground-state NI energy can be reached variationally, in principle
exactly, as follows,
\be\label{eq:exact_embedding_MF_ener}
E_0^{\rm
NI}=\min_{\bv}\left\{E^{\rm NI}_{\calC}[\bv]+E^{\rm
NI}_{\calE}[\bv]\right\},
\ee
where
\be\label{eq:min_MF_ener_cluster}
E^{\rm NI}_{\calC}[\bv]=\min_{\Phi^\calC}\mel{\Phi^\calC}{\hat{h}^{\calC}[\bv]}{\Phi^\calC}
\ee
and
\be\label{eq:min_MF_ener_env}
E^{\rm NI}_{\calE}[\bv]=2\min_{\tilde{\gamma}^{\calE}}\Tr\left[\tilde{\bm
h}^\calE[\bv]\tilde{\gamma}^{\calE}\right].
\ee
Dependencies in the Householder vector $\bv$ have been introduced, for clarity.
In order to evaluate the per-site energy [see
Eq.~(\ref{eq:MF_ener})], we can switch to the Householder representation,
\be
\begin{split}
\gamma_{10}&=\left[\bP\tilde{\bg}\bP\right]_{10}
=
\sum_{ij}P_{1i}\tilde{\gamma}_{ij}P_{j0}
\\
&=
\sum_{i}P_{1i}\tilde{\gamma}_{i0}
=
\sum^1_{i=0}P_{1i}\tilde{\gamma}_{i0}
\\
&=
\sum^1_{i,j=0}P_{1i}\tilde{\gamma}_{ij}P_{j0}
\\
&=\sum_{ij}P_{1i}\mel{\Phi^\calC}{\ddop_{i\sigma}\dop_{j\sigma}}{\Phi^\calC}P_{j0}
\\
&=\expval{\cdop_{1\sigma}\cop_{0\sigma}}_{\Phi^\calC},
\end{split}
\ee
thus leading to the final expression
\be\label{eq:final_eHF_from_PhiC}
e^{\rm
NI}(n)&=-4t\expval{\cdop_{1\sigma}\cop_{0\sigma}}_{\Phi^\calC}.
\ee
According to Eq.~(\ref{eq:final_eHF_from_PhiC}), the per-site NI energy can be evaluated
directly from the cluster, which is obviously a huge
simplification of the full-size problem. The exact NI per-site energy is
recovered when 
the minimizing Householder vector $\bv$ in
Eq.~(\ref{eq:exact_embedding_MF_ener}) is employed, thus providing the
optimal bath orbital. As readily seen from the latter
equation, the Householder vector connects the cluster to its environment
energy wise. In the present formalism, the optimal cluster (or,
equivalently, the optimal bath) minimizes the
sum of the cluster and environment energies. At the NI level, 
Eqs.~(\ref{eq:exact_embedding_MF_ener})--(\ref{eq:min_MF_ener_env})
are much more complicated than Eq.~(\ref{eq:RDM_from_occupied_orbs}) implementation wise, especially
because the Hamiltonian of the environment should in principle be
diagonalized for each trial Householder vector. However, once the full-size NI problem
is solved [with Eq.~(\ref{eq:RDM_from_occupied_orbs})],
Eqs.~(\ref{eq:exact_embedding_MF_ener})--(\ref{eq:min_MF_ener_env})
can be used for describing two-electron interactions. In the simplest embedding scheme,
which is described in the present work, 
electron correlation is introduced within the cluster while freezing the Householder vector to
its NI value. The embedding might then be systematically improved by $(i)$ updating
the Householder vector variationally, $(ii)$ describing correlations between the
cluster and the environment and, ultimately, $(iii)$ describing
correlation within the environment. Such refinements are left for future work.


\subsection{Correlation on top of the non-interacting Householder embedding}\label{subsec:correlated_embedding}

By analogy with DMET~\cite{knizia2012density}, we now introduce the exact two-electron repulsion
operator within the Householder cluster. For simplicity, we keep on
using the Householder vector 
$\bv$ evaluated from the NI density matrix of Eq.~(\ref{eq:RDM_from_occupied_orbs}). First we need to rewrite the
on-site repulsion operator
in the Householder representation [see Eq.~(\ref{eq:inverse_transf_less_explicit})],
\be
U\sum^{L-1}_{i=0}\hat{n}_{i\uparrow}\hat{n}_{i\downarrow}=\sum_{jklm}\tilde{U}_{jklm}\ddop_{j\uparrow}\dop_{k\uparrow}\ddop_{l\downarrow}\dop_{m\downarrow},
\ee      
where
\be\label{eq:2e_ints_HHrep}
\tilde{U}_{jklm}=U\sum^{L-                                                                                                                                                 1}_{i=0}
P_{ij}P_{ik}P_{il}P_{im}.
\ee 
After projecting onto the cluster, we obtain the following expression
for the interacting cluster Hamiltonian:
\be\label{eq:emb_Hamilt_int_bath}
\begin{split}
\hat{\mathcal{H}}^\calC&=\hat{h}^{\calC}+\sum^1_{j,k,l,m=0}\tilde{U}_{jklm}\ddop_{j\uparrow}\dop_{k\uparrow}\ddop_{l\downarrow}\dop_{m\downarrow}
\\
&=\hat{h}^{\calC}+U\ddop_{0\uparrow}\dop_{0\uparrow}\ddop_{0\downarrow}\dop_{0\downarrow}+\tilde{U}_1\ddop_{1\uparrow}\dop_{1\uparrow}\ddop_{1\downarrow}\dop_{1\downarrow},
\end{split}
\ee
where, according to Eq.~(\ref{eq:2e_ints_HHrep}), the effective interaction in the bath is determined from (the 
NI approximation to) the full lattice as follows,
\be\label{eq:interaction_bath}
\begin{split}
\tilde{U}_1&=U\sum^{L-1}_{i=1}P^4_{i1}
\\
&=U\left(1-8v_1^2+24v_1^4-32v_1^6+16v_1^4\sum^{L-1}_{i=1}v_i^4\right).
\end{split}
\ee
In the DMET terminology~\cite{bulik2014density,wouters2016practical},
using the complete embedding Hamiltonian of
Eq.~(\ref{eq:emb_Hamilt_int_bath}) is referred to as {\it interacting
bath} (IB) embedding. In this case, the interaction $\tilde{U}_1$ on the
bath site is taken into account. We should stress that, in the present
work, like in regular DMET, the bath orbital (which is described by the
operators
$\ddop_{1\sigma}$ and $\dop_{1\sigma}$) is determined from the
density matrix of the full {\it non-interacting} lattice, according to
Eq.~(\ref{eq:bath_orbital}). Therefore, the acronym IB should not be 
confused with the level of calculation of the bath orbital.   
In the {\it non-interacting bath} (NIB)
formulation~\cite{bulik2014density,wouters2016practical}, which is
commonly used in DMET and is also tested in Sec.~\ref{sec:results}, the interaction $\tilde{U}_1$
on the bath site is simply neglected. Note that, in the context of DMET, correlating the bath would require
the
computation of a correlated many-body wave function for the full
system~\cite{hermes2019multiconfigurational} or the introduction of frequency dependencies into the
theory~\cite{PRB21_Booth_effective_dynamics_static_embedding}. In the
present density matrix functional formulation,
correlation might be introduced
into the bath simply by employing in the Householder transformation [see
Eqs.~(\ref{eq:Pij_exp})--(\ref{eq:xi_final_exp})] a
correlated full-system-size density matrix. 
This can be
achieved at a reasonable computational cost~\cite{PCCP16_Giesbertz_avoiding-4index_transf_RDMFT} with
low-level natural orbital
functionals
(NOFs). Their (sometimes poor 
~\cite{mitxelena2017performance,mitxelena2017performance_corrigendum})
description 
of strongly correlated
energies may then be improved {\it via} the embedding, thus avoiding the
use of more sophisticated NOFs which can be more difficult to converge. 
In this context, we may actually proceed with successive Householder
transformations since the Householder
``impurity+bath'' cluster will in principle not be disconnected anymore
from its
environment, as discussed in Sec.~\ref{subsec:HHt_dens_mat}.
Applying a {\it second} Householder transformation to the
``bath+buffer+environment'' block of the Householder transformed density
matrix (see Fig.~\ref{fig:schematics_HT_1RDM}) would generate a second bath orbital. The interacting lattice
problem can then be projected onto the enlarged ``impurity+two bath
orbitals'' cluster. Applying further Householder transformations would
generate more bath orbitals and, ultimately, make the embedding
exact (because equivalent to the full lattice diagonalization
problem). Interestingly, in EwDMET, the enlarged number of bath orbitals is determined from the
order to which fragment spectral moments should be reproduced. A formal
connection between Ht-DMFET and EwDMET might be established at this
level.  
We leave the development of such an extension 
for future work. 
Note that, within a hybrid NOF/Householder scheme, we
will still be able to choose between IB and NIB formulations. 
In the rest of the paper, we employ an {\it uncorrelated} bath
(which is identical to the 
non-interacting bath for uniform systems).\\

Let us return to the embedding Hamiltonian in Eq.~(\ref{eq:emb_Hamilt_int_bath}). As the impurity occupation
may deviate from the lattice filling $n$ when solving the interacting
problem within the cluster, we introduce and
adjust a
chemical potential $\tilde{\mu}^{\rm imp}$ on the impurity site, in
complete analogy with DMET~\cite{wouters2016practical}, such that the
(two-electron) cluster
wave function 
\be\label{eq:PsiC_argmin}
\Psi^\calC=\argmin_{\Psi}\mel{\Psi}{\hat{\mathcal{H}}^\calC-\tilde{\mu}^{\rm
imp}\sum_{\sigma}\ddop_{0\sigma}\dop_{0\sigma}}{\Psi}
\ee
reproduces the desired occupation $n$, \ie
\be\label{eq:imp_dens_constraint}
\mel{\Psi^\calC}{\sum_{\sigma}\ddop_{0\sigma}\dop_{0\sigma}}{\Psi^\calC}\overset{!}{=}n.
\ee
Once the constraint in Eq.~(\ref{eq:imp_dens_constraint}) is fulfilled,
we obtain an approximate correlated expression for the per-site energy,
\be\label{eq:Ht-dmfet_persite_ener}
e(n)\approx
-4t\expval{\cdop_{1\sigma}\cop_{0\sigma}}_{\Psi^\calC}+U\expval{
\hat{n}_{0\uparrow}\hat{n}_{0\downarrow}
}_{\Psi^\calC},
\ee
where the density matrix element can be evaluated as follows [see
Eq.~(\ref{eq:inverse_transf})], 
\be
\begin{split}
\expval{\cdop_{1\sigma}\cop_{0\sigma}}_{\Psi^\calC}&=\left(1-2v^2_1\right)\expval{\ddop_{1\sigma}\dop_{0\sigma}}_{\Psi^\calC}
\\
&=\left(1-2v^2_1\right)\tilde{\gamma}^{\Psi^\calC}_{10},
\end{split}
\ee
and the impurity double occupation simply reads $\expval{
\hat{n}_{0\uparrow}\hat{n}_{0\downarrow}
}_{\Psi^\calC}=\expval{\ddop_{0\uparrow}\dop_{0\uparrow}\ddop_{0\downarrow}\dop_{0\downarrow}}_{\Psi^\calC}=d^{\Psi^\calC}_0$.
Note that the Hamiltonian in Eq.~(\ref{eq:PsiC_argmin}) describes an
asymmetric Hubbard dimer whose 
two-electron {\it singlet}
ground-state energy can be expressed 
analytically in terms of $\tilde{t}$, $U$, $\tilde{U}_1$,
$\tilde{\varepsilon}_1$, and $\tilde{\mu}^{\rm imp}$. For that purpose,
we simply need to shift the local one-electron potential (on both
impurity and bath sites) by the ``constant''
$\left(\tilde{\mu}^{\rm
imp}-\tilde{\varepsilon}_1\right)/2$ 
{\it and} to symmetrize
the dimer interaction
wise~\cite{senjean2017local}, 
\be\label{eq:symm_int_dimer}
\begin{split}
&\hat{\mathcal{H}}^\calC-\tilde{\mu}^{\rm
imp}\sum_{\sigma}\ddop_{0\sigma}\dop_{0\sigma}
\rightarrow 
\sum_\sigma
-\tilde{t}
\left(\ddop_{0\sigma}\dop_{1\sigma}+\Hc\right)
\\
&
+U_{\rm
eff}\sum^1_{i=0}\ddop_{i\uparrow}\dop_{i\uparrow}\ddop_{i\downarrow}\dop_{i\downarrow}
+\dfrac{\Delta v_{\rm
eff}}{2}\sum_{\sigma}\left(\ddop_{1\sigma}\dop_{1\sigma}-\ddop_{0\sigma}\dop_{0\sigma}\right),
\end{split}
\ee
where the effective interaction and potential read 
\be\label{eq:Ueff_exp}
U_{\rm
eff}=\dfrac{U+\tilde{U}_1}{2}
\ee
and
\be\label{eq:Deltav_eff_exp}
\Delta v_{\rm
eff}=
\tilde{\mu}^{\rm
imp}+\tilde{\varepsilon}_1
+
\dfrac{\tilde{U}_1-U}{2}
,
\ee 
respectively. Note that the Hamiltonians in the left- and right-hand
side of Eq.~(\ref{eq:symm_int_dimer}) share the same ground-state wave
function $\Psi^\calC$~\cite{senjean2017local}. The corresponding ground-state energies are
immediately deduced from the analytical expression given in
Ref.~\onlinecite{carrascal2015hubbard}.
The chemical potential $\tilde{\mu}^{\rm imp}$ that fulfills Eq.~(\ref{eq:imp_dens_constraint}) or, equivalently, the
effective potential $\Delta v_{\rm
eff}$, can be determined straightforwardly, for
example, by Lieb maximization~\cite{senjean2017local}. Both $\tilde{\gamma}^{\Psi^\calC}_{10}$ and $d^{\Psi^\calC}_0$ can
then be evaluated analytically {\it via} the Hellmann--Feynman
theorem (see Appendix A in Ref.~\onlinecite{senjean2019projected}).
In the DMET terminology~\cite{wouters2016practical}, 
Eqs.~(\ref{eq:PsiC_argmin})--(\ref{eq:Ht-dmfet_persite_ener}) describe a {\it
single-shot} embedding.

\subsection{Summary of the present implementation and connection to 
DMET}\label{subsec:summary_connections}

We have described the single-shot embedding of a single impurity site in the
particular case of a 1D Hubbard lattice. The full procedure can be
summarized as follows. First we solve the non-interacting problem. In
the present work we diagonalize the bare hopping matrix
[see Eq.~(\ref{eq:hopping_matrix})] and
construct the (idempotent) ground-state density matrix of the full
lattice for a  
given and fixed number $N=nL$ of electrons. The latter density matrix gives immediately
access to the
bath orbital thanks to the Householder transformation [see the
expression in Eq.~(\ref{eq:bath_orbital}); see also
Eqs.~(\ref{eq:Householder_vec}) and (\ref{eq:xi_final_exp})]. Then we
project the original interacting lattice Hamiltonian of
Eq.~(\ref{eq:uniform_hamilt_1D}) onto the ``impurity+bath'' many-body
subspace, which gives the cluster Hamiltonian expression of
Eq.~(\ref{eq:emb_Hamilt_int_bath}). At this level, we can decide to keep
the interaction in the bath (IB formulation) or to remove it (NIB
formulation). Finally, a chemical potential is introduced and adjusted
on the embedded impurity to ensure that its occupation matches the
lattice filling $n$ [see Eqs.~(\ref{eq:PsiC_argmin}) and
(\ref{eq:imp_dens_constraint})]. A correlated per-site energy can then be evaluated
from the (interacting) cluster many-body wave function [see
Eq.~(\ref{eq:Ht-dmfet_persite_ener})]. For analysis purposes, 
NIB results obtained with two or three
impurities are presented
in Sec.~\ref{sec:results}. In the latter case, we used a block Householder
transformation~\cite{AML99_Rotella_Block_Householder_transf} where the
column vector $\bX-\bY$ in Eq.~(\ref{eq:HH_vec}) is replaced by a $L\times N_{\rm imp}$ matrix
which is denoted $\bV$ in Ref.~\onlinecite{AML99_Rotella_Block_Householder_transf},
$N_{\rm imp}$ being the number of impurities. The Householder
transformation matrix is then modified as follows, 
\be\label{eq:multiple_imp_HH_transf}
\bP=\bI-2\bv\bv^\dagger\rightarrow
\bI-2\bV\left[\bV^\dagger\bV\right]^{-1}\bV^\dagger.
\ee
A detailed derivation of multiple-impurity Ht-DMFET will be presented
in a separate work.\\      

Let us stress that, at a given level of approximation (we performed a
{\it single-shot} embedding for simplicity but stronger mapping
constraints could of course be employed), the IB formulations of Ht-DMFET and standard DMET are formally
equivalent. Indeed, for a non-interacting (or mean-field) lattice, the
bath orbitals constructed from the Householder transformation and the
Schmidt decomposition are identical, as shown in
Sec.~\ref{subsec:compar_Schmidt_decomp}. As a result, when projecting
the lattice interactions onto the ``impurity+bath'' cluster, both
approaches will lead to the same
embedding Hamiltonian [the one in Eq.~(\ref{eq:emb_Hamilt_int_bath})].
More generally, as long as the bath is {\it uncorrelated}, which means
that we describe the full lattice with an idempotent (non-interacting or
mean-field) density matrix, Ht-DMFET becomes a
simpler to implement (but equivalent) version of DMET. Note also that
the
Householder transformation can in principle be substituted for the
Schmidt decomposition when constructing hole and particle bath
states in EwDMET [see Eqs.~(25) and (26) in
Ref.~\onlinecite{JCP19_Booth_Ew-DMET_hydrogen_chain}]. In this case, we
would need to construct density matrices for cationic and anionic
systems [see Eq.~(10) in Ref.~\onlinecite{fertitta2018rigorous}].\\  

Finally, if we just see in the Householder transformation a simplification of the
Schmidt decomposition in the particular case of non-interacting systems, it becomes clear
that, by complete analogy with DMET, 
Ht-DMFET can be extended to more general Hamiltonians like, for example, quantum chemical ones
written in a localized orbital basis. In
this case, several fragments (each consisting of multiple localized
impurity orbitals) would be employed. More precisely, once a mean-field density
matrix has been computed for the full molecule, it can be Householder
transformed with respect to a given fragment, thus providing a set of
bath orbitals from which correlated local properties (on the fragment) can be
evaluated. Each fragment will have its own Householder transformation
but all will apply to the same (full-system) density matrix. Obviously,
the block version of the Householder transformation~\cite{AML99_Rotella_Block_Householder_transf} would be employed in
this case. Moreover, as already mentioned in
Sec.~\ref{subsec:correlated_embedding}, the Householder transformation offers the possibility
to correlate the bath through the density matrix, a feature that DMET
does not have and from which we may benefit. Work is currently in
progress in this direction.  

\section{Results and discussion}\label{sec:results}

In this section, Ht-DMFET 
is applied to a large ($L=400$) uniform Hubbard
ring. The
hopping parameter is set to $t=1$. In order to remove pathological
degeneracies from the $N$-electron full-size non-interacting
calculation, periodic/antiperiodic boundary conditions are used when
$\frac{N}{2}$ is odd/even. Comparison is made with the Bethe Ansatz (BA) results which are exact in
the thermodynamic limit~\cite{NoMott_Hubbardmodel}. We note that 
the single-impurity DMET results presented in
Ref.~\onlinecite{knizia2012density} were obtained without the interaction in
the bath. The interacting-bath results presented in the following will be of particular interest in this
respect.
For analysis purposes, multiple-impurity Ht-DMFET calculations have been
performed at the NIB level [see Eq.~(\ref{eq:multiple_imp_HH_transf})]. In the latter case, the correlated embedded
impurity 
problem has been solved numerically 
at the {\it density matrix renormalization group} (DMRG) level with the Block code~\cite{chan2002highly,chan2004algorithm,ghosh2008orbital,sharma2012spin,guo2016n}.
\\
\begin{figure}
\begin{center}
\includegraphics[scale=0.5]{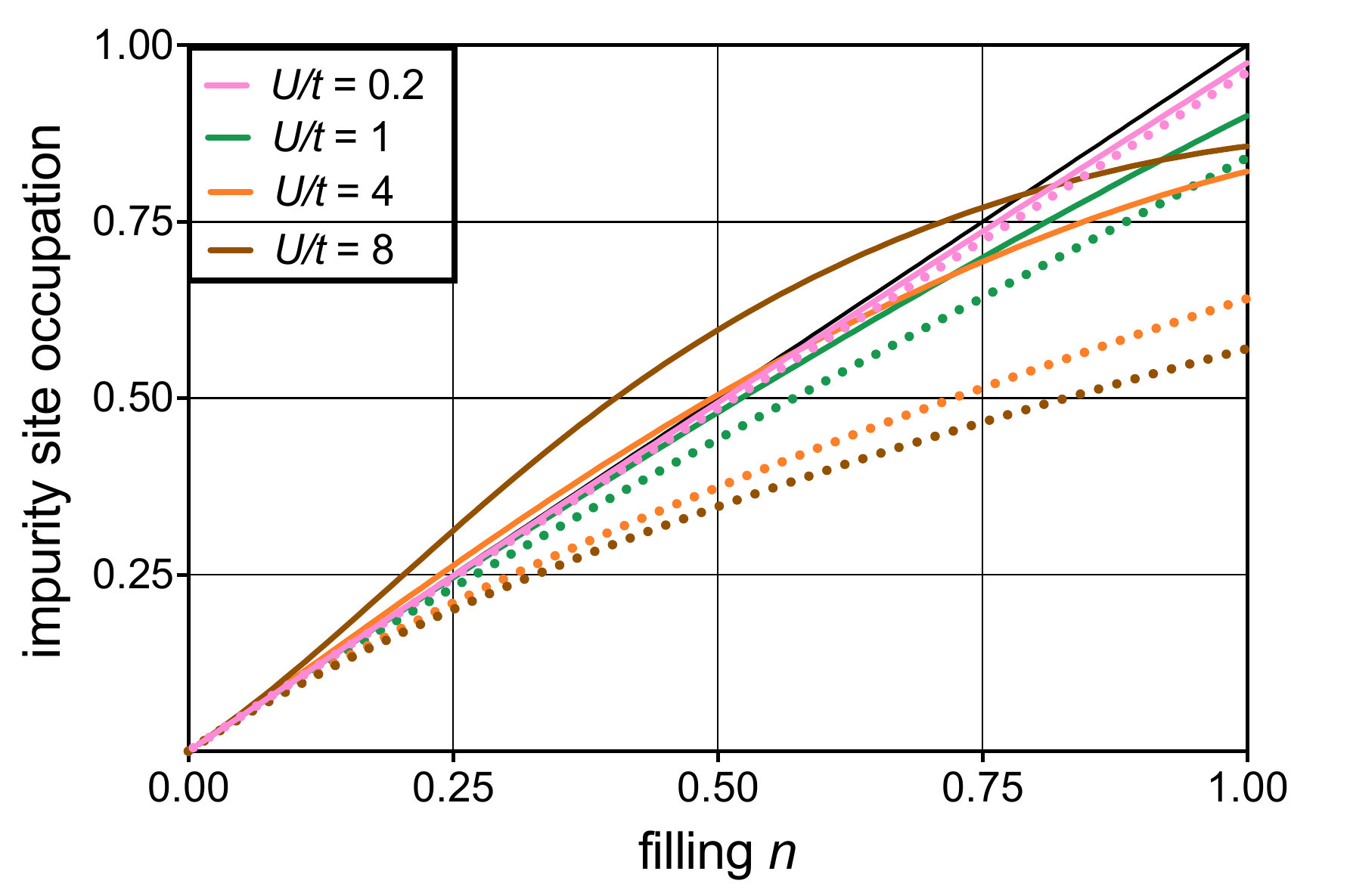}
\caption{Householder cluster's (single) impurity site occupation plotted 
as a
function of the lattice filling $n$ in
various correlation regimes for 
$\tilde{\mu}^{\rm imp}=0$ [see Eq.~(\ref{eq:PsiC_argmin})]. 
Both interacting (solid lines) and
non-interacting (dotted lines) bath cases are shown for analysis
purposes. The reference black straight line corresponds to the desired situation
(which is ultimately reached by adjusting $\tilde{\mu}^{\rm imp}$) where the embedded
impurity occupation matches the lattice filling.}
\label{Fig:deviations_in_dens_NIB_and_IB}
\end{center}
\end{figure}
\begin{figure}
\begin{center}
\includegraphics[scale=0.5]{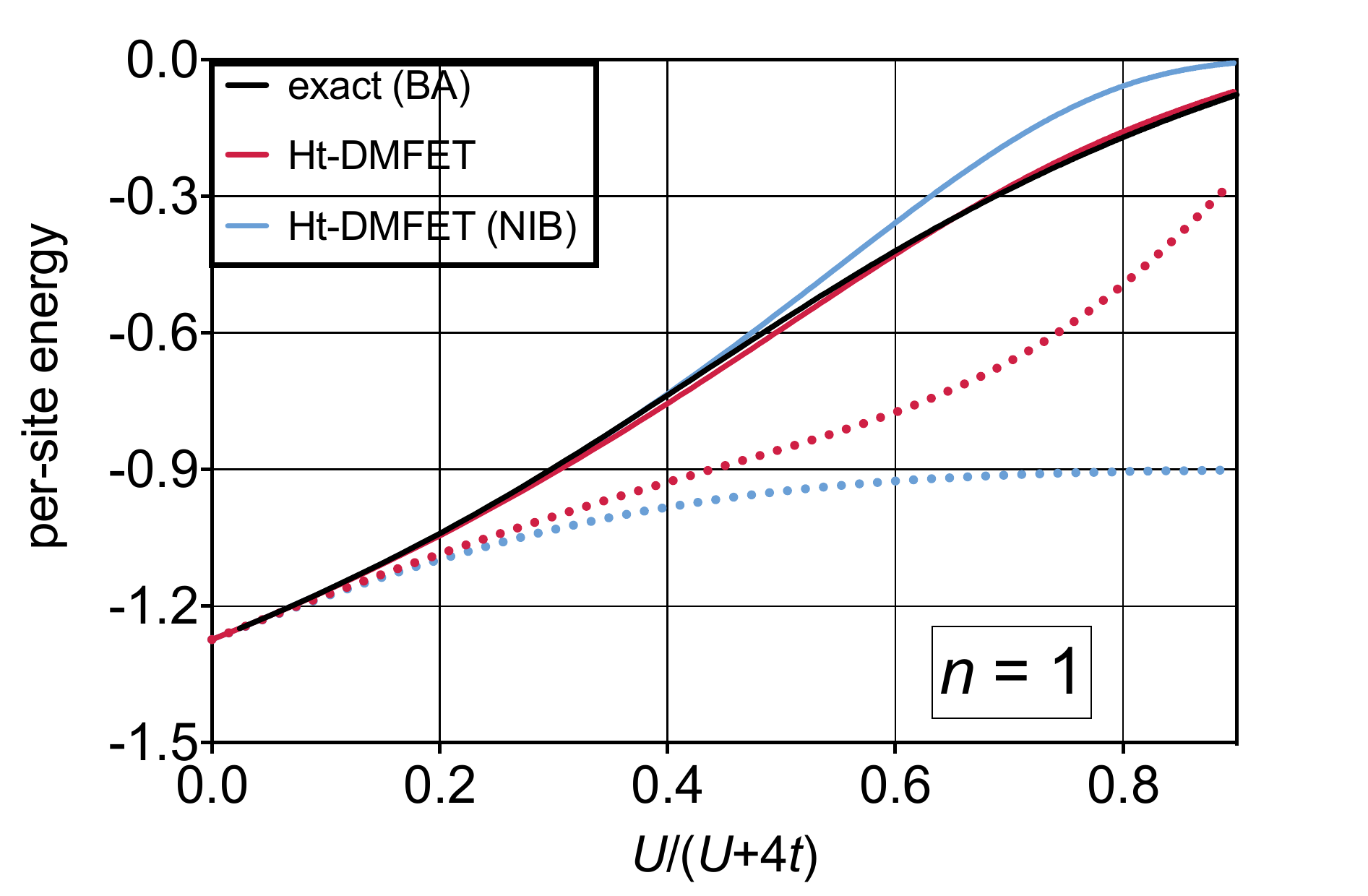}
\caption{(Single-impurity) Ht-DMFET per-site energy plotted as a function of the interaction
strength at half-filling. Results obtained for $\tilde{\mu}^{\rm imp}=0$
(dotted lines) and/or without interaction in the bath (blue lines) are
shown for analysis purposes. Comparison is made with the exact Bethe
Ansatz (BA) result.}
\label{Fig:per-site_energy_as_a_function_of_U}
\end{center}
\end{figure}
\begin{figure}
\begin{center}
\includegraphics[scale=0.5]{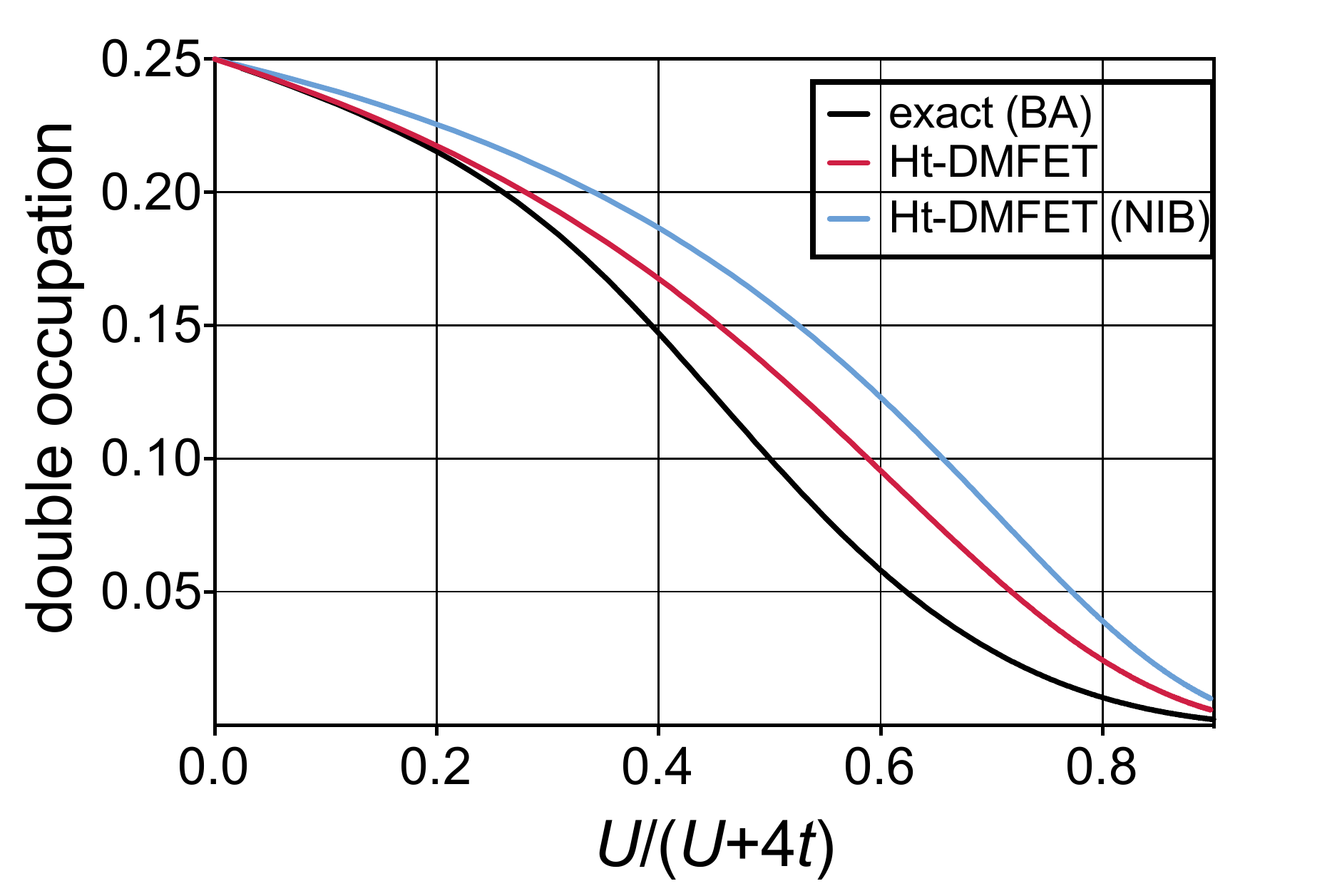}
\caption{Ht-DMFET (single) impurity site double occupation $\expval{
\hat{n}_{0\uparrow}\hat{n}_{0\downarrow}
}_{\Psi^\calC}$ plotted
as a function of the interaction strength at half-filling ($n=1$).
Comparison is made with the
exact Bethe Ansatz (BA) result. Non-interacting bath (NIB) results are shown for analysis purposes.}
\label{Fig:double_occupation_as_a_function_of_U}
\end{center}
\end{figure}

Let us first discuss the single-impurity results. We plot in
Fig.~\ref{Fig:deviations_in_dens_NIB_and_IB} the deviation in occupation
of the (two-electron) Householder cluster's impurity site 
from the filling $n$ of the lattice when no additional chemical potential $\tilde{\mu}^{\rm imp}$ is introduced.
As expected from Sec.~\ref{subsec:embedding_for_MF}, the correct occupation $n$ is recovered in
the non-interacting $U=0$ case (not shown). 
When $U>0$, we observe, in the NIB
formulation, a systematic
depletion of the impurity site as $U$ increases. For given $U$ and $n$
values, adding the ($U$- and $n$-dependent) interaction $\tilde{U}_1$
[see Eq.~(\ref{eq:interaction_bath})] on the bath site 
has the opposite effect. 
The fact that the interactions on the
impurity and the bath do not compensate each other occupation wise can
be understood as follows. Let us consider, for example, the half-filled
case ($n=1$) for which $\tilde{\varepsilon}_1=0$. When $\tilde{\mu}^{\rm
imp}=0$, in the NIB case, the
cluster is equivalent to a symmetric-in-interaction Hubbard dimer
with effective interaction $U_{\rm
eff}=U/2$ and potential $\Delta v_{\rm
eff}=-U/2$ [see Eqs.~(\ref{eq:Ueff_exp}) and (\ref{eq:Deltav_eff_exp})],
thus leading to $
\Delta v_{\rm
eff}
/U_{\rm
eff}\overset{\rm NIB}{=}-1
$, which obviously will not give the symmetric density profile expected
for $n=1$. As shown in Fig.~1 of Ref.~\onlinecite{deur2017exact}, the occupation of the impurity will decrease
with $U_{\rm
eff}$ (and therefore $U$) and tend to 0.5 as $U/t\rightarrow +\infty$,
which is in agreement with Fig.~\ref{Fig:deviations_in_dens_NIB_and_IB}.
In the IB case (see Sec.~\ref{subsec:correlated_embedding}), the situation is quite different since
\be
\dfrac{
\Delta v_{\rm
eff}
}{U_{\rm
eff}}\overset{\rm
IB}{=}\dfrac{\tilde{U}_1-U}{\tilde{U}_1+U}=\dfrac{
\displaystyle
\sum^{L-1}_{i=1}P^4_{i1}-1}{
\displaystyle
\sum^{L-1}_{i=1}P^4_{i1}+1},
\ee
thus leading to $\abs{\Delta v_{\rm
eff}/U_{\rm
eff}}<1$. In this respect, we are closer to a symmetric dimer, which is
an improvement. Nevertheless, since $\Delta v_{\rm
eff}\neq 0$, the correct occupation ($n=1$) will not be recovered, in
general. By computing $\tilde{U}_1$ numerically, we obtained $\tilde{U}_1\approx U/3$ at
half-filling, which gives $\Delta v_{\rm
eff}/U_{\rm
eff}\overset{\rm IB}{\approx}-1/2$. In this case, as $U$ (and therefore $U_{\rm
eff}$) starts deviating from 0, there will still be an electron depletion of the
impurity site [see the top right panel of Fig.~1 in
Ref.~\onlinecite{deur2017exact}]. Interestingly, for larger $U$ values, the impurity occupation will
start increasing and will approach (from below) the correct 1.0 value, as shown in the bottom panels of Fig.~1 in
Ref.~\onlinecite{deur2017exact}. This is again in perfect agreement with
the IB results of Fig.~\ref{Fig:deviations_in_dens_NIB_and_IB}. In summary, when no chemical potential
is introduced on the impurity, the IB and NIB impurity occupation
profiles differ substantially. This difference is driven by the effective ratio $\Delta v_{\rm
eff}/U_{\rm
eff}$. Despite this difference, the IB scheme cannot reproduce the 
correct occupation and, as a result, introducing a chemical potential
$\tilde{\mu}^{\rm imp}$ remains necessary, like in
DMET~\cite{wouters2016practical}.\\

Its importance in the calculation of
per-site energies is illustrated in
Fig.~\ref{Fig:per-site_energy_as_a_function_of_U} in the particular case
of a half-filled lattice. Once a proper $\tilde{\mu}^{\rm imp}$ value is
employed (which will be the case in the rest of the discussion), thus ensuring that the filling and the impurity occupation match, the
error becomes substantial in the
strongly correlated regime only if the interaction in the bath is
neglected. We note that, in the latter case, we reproduce the
single-impurity DMET results of Ref.~\onlinecite{knizia2012density},
as expected from the analysis in Sec.~\ref{subsec:compar_Schmidt_decomp}.
The agreement with the BA results is almost perfect in all correlation
regimes once the interaction in the bath is
restored. This success may be related to the fact that, like in our approximate Ht-DMFET
scheme, the true (correlated) Householder cluster contains exactly two
electrons in the half-filled case, as a consequence of the hole-particle
symmetry (see Appendix~\ref{appendix:Nc_half-filled_case}).
Nevertheless, even though the interaction in the bath improves on the
impurity double occupation, the error remains substantial when electron
correlation is strong, as shown in
Fig.~\ref{Fig:double_occupation_as_a_function_of_U}. In order to further reduce
the error, more impurities should be introduced into the
cluster~\cite{knizia2012density}. Therefore, the
success of the present (single-impurity) Ht-DMFET at half-filling relies
also on error cancellations in the  
evaluation of the (total) per-site energy.\\ 

\begin{figure}
\begin{center}
\includegraphics[scale=0.5]{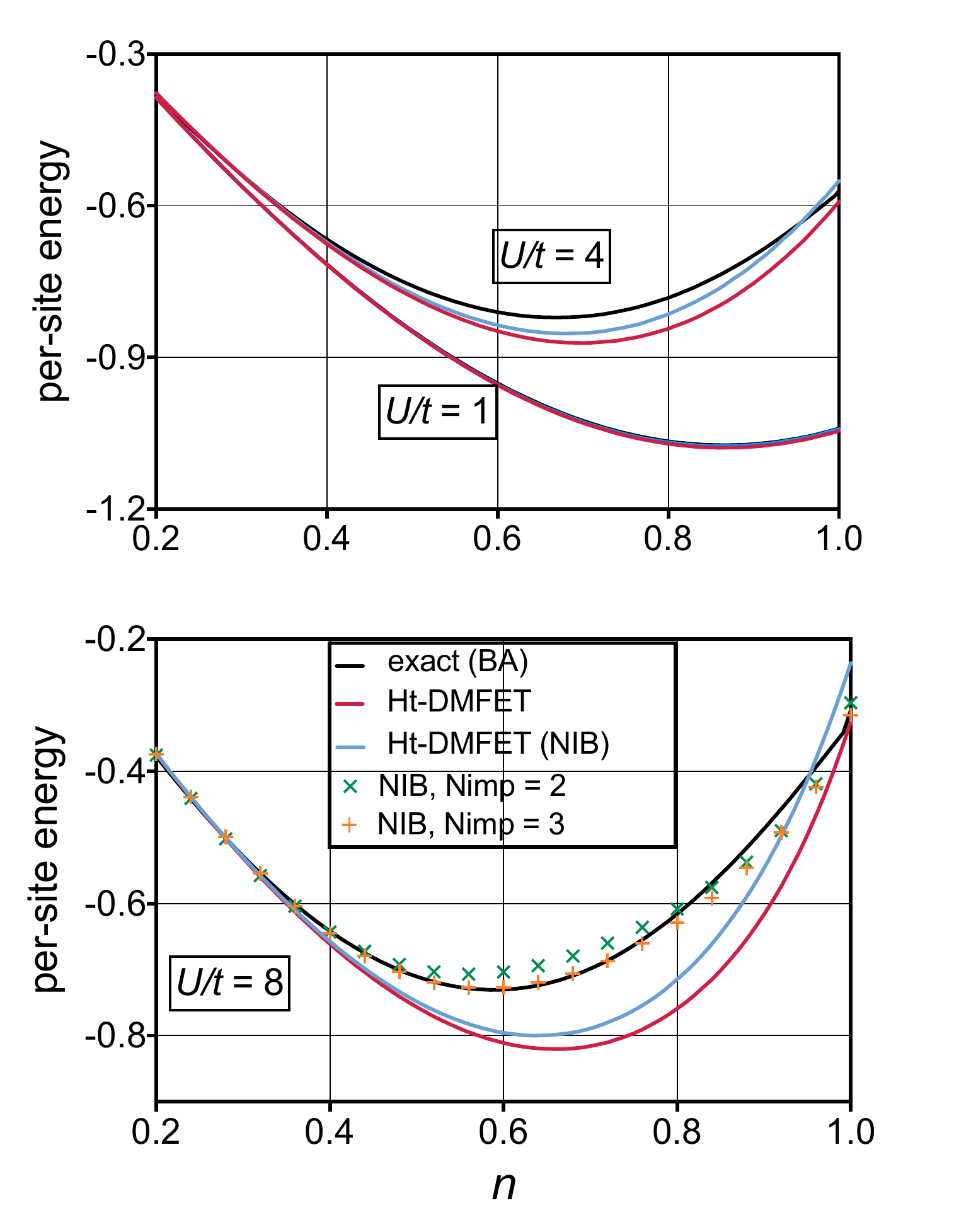}
\caption{
Ht-DMFET per-site energies plotted as a function of the lattice
filling $n$ for various correlation regimes. Results obtained with a
single impurity are shown as (colored) solid lines. The blue color
corresponds to the non-interacting bath (NIB) case. In the strongly correlated
$U/t=8$ regime (bottom panel), NIB results obtained
with two
($N_{\rm imp}=2$) and
three ($N_{\rm imp}=3$) impurities are also shown (as points), for analysis purposes (see Sec.~\ref{subsec:summary_connections} for
further details).
Comparison is made with the exact Bethe Ansatz (BA)
results (black solid lines). In the weakly $U/t=1$ correlated case (top panel), exact and
approximate results are almost indistinguishable.
}
\label{Fig:per-site_energy_as_a_function_of_n}
\end{center}
\end{figure}

Away from half-filling, the performance of (single-impurity) Ht-DMFET deteriorates as $U/t$
increases, as shown in
Fig.~\ref{Fig:per-site_energy_as_a_function_of_n}, probably because
fluctuations in the number of electrons within our (``single
impurity+single bath'') cluster are not allowed in our approximate embedding. As discussed in
Sec.~\ref{subsec:HHt_dens_mat}, away from half-filling, the cluster becomes an open subsystem 
as soon as $U/t$ deviates from
zero. Surprizingly, in this density regime, per-site energies are in
better agreement with the BA values when the interaction in the bath is
neglected. Again, in the latter case, we recover the single-impurity DMET
results of Ref.~\onlinecite{knizia2012density}. 
As expected~\cite{knizia2012density,bulik2014density}
and shown in the bottom panel of Fig.~\ref{Fig:per-site_energy_as_a_function_of_n},
the results dramatically improve when a larger fragment (consisting of
two or three impurities) is embedded, even at the simplest NIB level
of approximation.
\\     

\begin{figure}
\begin{center}
\includegraphics[scale=0.5]{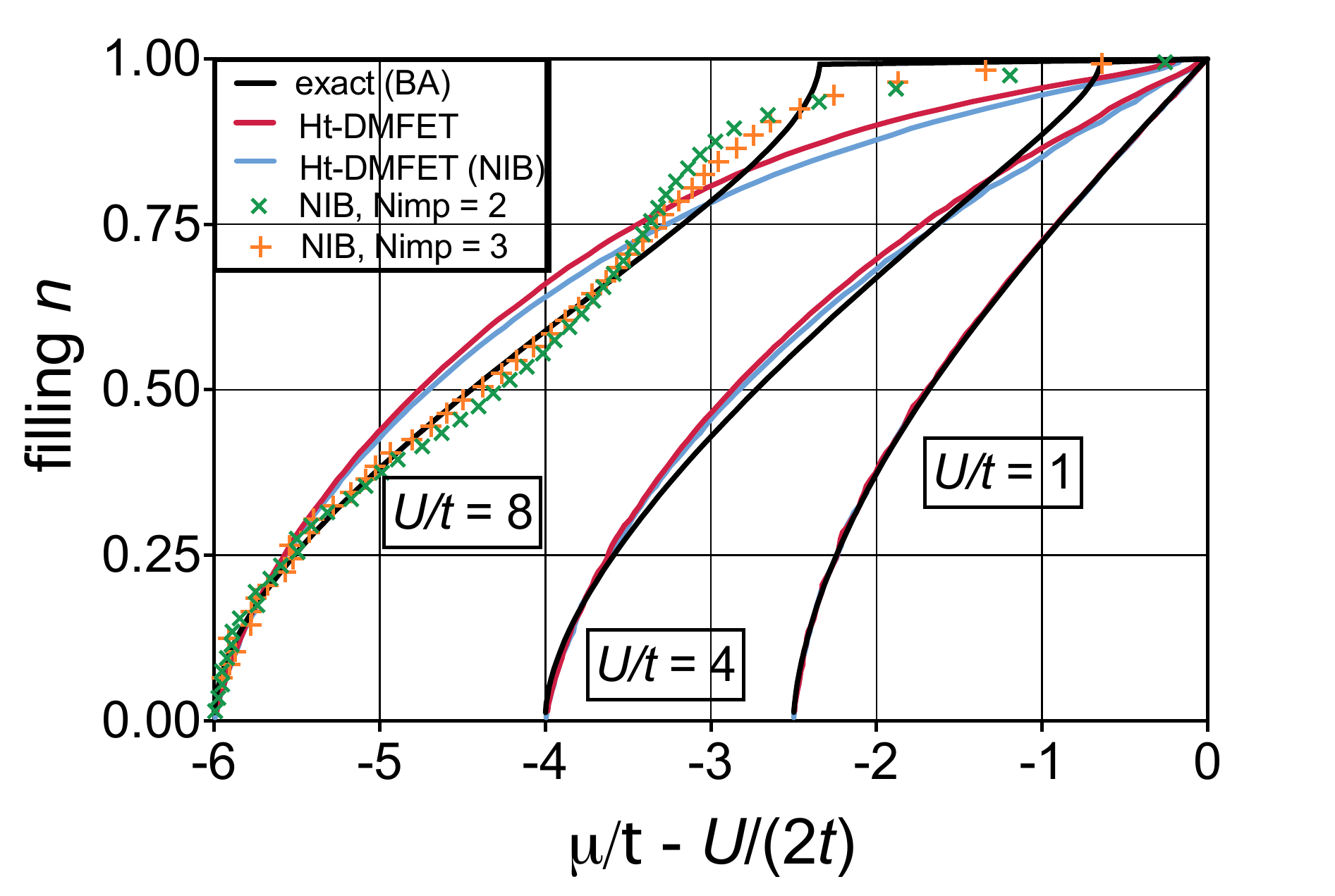}
\caption{
Lattice filling plotted, {\it via} the relation $\mu\equiv
\mu(n)=\partial e(n)/\partial n$, as a function of the (lattice) chemical potential
$\mu$ at the
Ht-DMFET level of calculation for various correlation regimes. (Single-impurity) non-interacting bath (NIB) results are 
shown as solid blue lines. In the strongly correlated $U/t=8$ case, NIB
results obtained with $N_{\rm imp}=2$ and $N_{\rm imp}=3$
impurities are also shown (as points), for analysis purposes.
Comparison is made with the
exact Bethe Ansatz (BA) results. 
}
\label{density-driven_Mott-Hubbard}
\end{center}
\end{figure}

Finally, we investigate in Fig.~\ref{density-driven_Mott-Hubbard} the
density-driven Mott--Hubbard transition {\it via} the evaluation of the density-functional $\mu(n)=\partial e(n)/\partial n$ chemical potential
from the Ht-DMFET energy expression of Eq.~(\ref{eq:Ht-dmfet_persite_ener}). 
As expected from
Ref.~\onlinecite{knizia2012density}, at the single-impurity level, there is no gap opening when the
interaction in the bath is neglected. Restoring the interaction in the
bath has actually no impact on the transition. 
In the light of
Sec.~\ref{subsec:HHt_dens_mat}, we can reasonably assume that Ht-DMFET fails in this
case because it relies on a closed two-electron ``single impurity+single
bath'' cluster. Already
at the NIB level of approximation, 
the embedding of a larger fragment (consisting of two or three
impurities) substantially improves the results. Nevertheless, even in
this case, the gap remains closed, which is in perfect agreement with
the DET results of Ref.~\onlinecite{bulik2014density}. As we perform 
single-shot embeddings (where we only require the embedded impurity to
reproduce the correct filling $n$), we expect from Ref.~\onlinecite{bulik2014density}
the transition to be better described at the multiple-impurity level
when the interactions in the bath are taken into account. It would also
be interesting to see how Ht-DMFET performs when a
correlated (through the density matrix) bath is employed. This is left for future
work.

\section{Conclusions and perspectives}\label{sec:conclusions}

Similar in spirit to DMET, a (static and zero-temperature) single-impurity Householder transformed density matrix functional embedding theory
(Ht-DMFET) has been derived. The theory has been applied to the 1D
Hubbard model. In the
non-interacting case, the formal reduction of the full lattice to a
two-electron dimer is exact. Thanks to the Householder transformation, the bath site
can be determined (analytically) from
the density matrix of the (full) lattice. Alternatively,
one may determine, in principle exactly, the Householder vector $\bv$ (which defines the
transformation) by minimizing the sum of the $\bv$-dependent Householder cluster and
environment energies. While the two-site ``impurity+bath''
cluster problem is trivially solved, the ground-state energy of the
cluster's environment must be evaluated for each trial 
vector $\bv$. Even though such a strategy is uselessly complicated in practice, because the
non-interacting full-size problem can be solved directly, it is enlightening in many
ways. 
First, it clearly shows that the optimal cluster cannot be determined
without learning from its environment
(and therefore from the full lattice). The two subsystems
``communicate'' through the
Householder vector. 
Secondly, the resulting {\it variational} character of the bath might be exploited when electron correlation is introduced into the
theory. One could also extract an approximate Householder
vector from many-body perturbation theory. This is left for future work.   
Finally, the single occupied spin-orbital that overlaps with the impurity
is, from the beginning, automatically separated from the core embedding
orbitals which, in the terminology of Ht-DMFET, belong to the Householder environment.
As a result, for non-interacting systems, Ht-DMFET is equivalent to  
DMET. It is however simpler since the embedding is constructed analytically from the
lattice representation of the density matrix.\\        

Starting from the exact non-interacting and {\it closed} two-electron Householder cluster, correlation
can be introduced straightforwardly by Householder transforming (and projecting onto the
cluster) the on-site two-electron repulsion operator, which is defined
in the lattice representation. 
At half-filling ($n=1$), the resulting (approximate) Ht-DMFET 
per-site energies are in almost perfect agreement with the Bethe
Ansatz (BA) results in all correlation regimes provided that 
{\it(i)} a chemical potential is introduced on the
impurity site, like in DMET, thus ensuring that the correct
filling is reproduced, and {\it(ii)} the interaction in the bath is
taken into account. 
The good performance of Ht-DMFET in this case can be partly related to the
fact that, at half-filling, the true (correlated) Householder cluster
contains exactly two electrons, as a
consequence of the hole-particle symmetry.
Away from half-filling, the deviation from the BA results 
becomes substantial in the strongly correlated regime.
The results
dramatically improve when a larger fragment (consisting of two or three
impurities) is embedded. In the latter case, a block Householder
transformation was
employed~\cite{AML99_Rotella_Block_Householder_transf}.
The failure of the single-impurity embedding away from half-filling may originate
from the fact that, unlike in our approximate Ht-DMFET, the number of
electrons in the exact
(correlated) Householder cluster can be fractional. This is probably the reason why, like single-impurity DMET with a
non-interacting bath~\cite{knizia2012density}, (single-impurity) Ht-DMFET
fails in 
describing the density-driven Mott--Hubbard transition, whether the bath
is interacting or not.\\ 
 
In the light of recent advances in DMET and related approaches, several
extensions of the present work can already be foreseen. A
multiple-impurity implementation of Ht-DMFET that is applicable to more
general Hamiltonians (like the quantum chemical ones) is the most urgent. Preserving
a Householder cluster that is disconnected from its environment is convenient in
practice but only relevant in the
non-interacting case. Applying the Householder transformation to the
Kohn--Sham density matrix would make the approach formally exact.
Introducing site (or orbital) occupation mapping constraints in a
self-consistency loop, like in
DET~\cite{bulik2014density} or SDE~\cite{mordovina2019self}, would make
complete sense in this context. Work is currently in progress in these
directions.

\section*{Acknowledgments}

The authors would like to thank Laurent Mazouin, Bruno Senjean, and
Mauricio Rodriguez-Mayorga for fruitful
discussions. They also thank LabEx CSC (ANR-10-LABX-0026-CSC) and ANR
(ANR-19-CE29-0002 DESCARTES project) for funding. 

\clearpage

\begin{appendices}
\appendix
\numberwithin{equation}{section}
\setcounter{equation}{0}

\section{Derivation of the Householder transformation} 
\label{appendix:YeqPX}

We start from the decomposition
\be\label{eq:Ydecomp}
\bY=(\bI-\bv\bv^\dagger)\bY+\bv\bv^\dagger\bY.
\ee
Using real algebra and the constraint $\abs{\bX}=\abs{\bY}$ gives, according to Eq.~(\ref{eq:HH_vec}),
\be\label{eq:simp_proj_Y}
\begin{split}
\bv^\dagger\bY&=\dfrac{\bX^\dagger\bY-\abs{\bY}^2}{\abs{\bX-\bY}}
\\
&=\dfrac{\bY^\dagger\bX-\abs{\bX}^2}{\abs{\bX-\bY}}
\\
&=\dfrac{\left(\bY^\dagger-\bX^\dagger\right)\bX}{\abs{\bX-\bY}}
\\
&=-\bv^\dagger\bX,
\end{split}
\ee
and
\be
\left(\bI-\bv\bv^\dagger\right)\bv=\bv-\left(\bv^\dagger\bv\right)\bv=0,
\ee
or, equivalently, 
\be\label{eq:proj_hyperplane}
\left(\bI-\bv\bv^\dagger\right)\bX=\left(\bI-\bv\bv^\dagger\right)\bY.
\ee
Combining Eqs.~(\ref{eq:Ydecomp}), (\ref{eq:simp_proj_Y}), and
(\ref{eq:proj_hyperplane}) leads to
\be
\bY=\left(\bI-\bv\bv^\dagger\right)\bX-\bv\bv^\dagger\bX\equiv \bP\bX,
\ee
where $\bP=\bI-2{\bv}{\bv}^\dagger$. 
%
\section{Exact properties of the Householder cluster for a 
half-filled 1D interacting lattice}\label{appendix:Nc_half-filled_case}

Let us evaluate the matrix element 
\be
\left[\tilde{\bm\gamma}^2\right]_{j0}=\left[\bP{\bm\gamma}^2\bP\right]_{j0}=\sum_{klm}P_{jk}\gamma_{kl}\gamma_{lm}P_{m0}
,
\ee
or, equivalently,
\be
\begin{split}
\left[\tilde{\bm\gamma}^2\right]_{j0}&=
\sum_{kl}P_{jk}\gamma_{kl}\gamma_{l0}\\
&\overset{j>0}{=}\sum_{l}P_{j1}\gamma_{1l}\gamma_{l0}+\sum_{k>1}\sum_lP_{jk}\gamma_{kl}\gamma_{l0}
\\
&\overset{j>0}{=}P_{j1}\gamma_{10}\left(\gamma_{00}+\gamma_{11}\right)+P_{j1}\sum_{l>1}\gamma_{1l}\gamma_{l0}
\\
&\quad+\gamma_{00}\sum_{k>1}P_{jk}\gamma_{k0}+\gamma_{10}\sum_{k>1}P_{jk}\gamma_{k1}
\\
&\quad\quad+\sum_{k>1}\sum_{l>1}P_{jk}\gamma_{kl}\gamma_{l0}.
\end{split}
\ee
Since $\gamma_{ii}=n_{\sigma}$ is the uniform
spin occupation in the system and $n=2n_{\sigma}$ is the lattice filling, it
comes
\be
\begin{split}
\left[\tilde{\bm\gamma}^2\right]_{j0}&\overset{j>0}{=}nP_{j1}\gamma_{10}+
n\sum_{k>1}P_{jk}\gamma_{k0}
\\
&\quad+
P_{j1}\sum_{l>1}\gamma_{1l}\gamma_{l0}
+\gamma_{10}\sum_{k>1}\delta_{jk}\gamma_{k1}
\\
&\quad-\dfrac{2v_j\gamma_{10}}{\sqrt{2\xi\left(\xi-\gamma_{10}\right)}}\sum_{k>1}\gamma_{k0}\gamma_{k1}
\\
&\quad\quad
+\sum_{k>1}\sum_{1<l\neq k}\delta_{jk}\gamma_{kl}\gamma_{l0}
\\
&\quad
-\dfrac{2v_j}{\sqrt{2\xi\left(\xi-\gamma_{10}\right)}}\sum_{k>1}\sum_{1<l\neq k}\gamma_{k0}\gamma_{kl}\gamma_{l0}
.
\end{split}
\ee 
Moreover,
\be
\begin{split}
P_{j1}\gamma_{10}+\sum_{k>1}P_{jk}\gamma_{k0}
&=\delta_{j1}\gamma_{10}-2v_j\dfrac{(\gamma_{10}-\xi)\gamma_{10}}{\sqrt{2\xi\left(\xi-\gamma_{10}\right)}}
\\
&\quad-2v_j\sum_{k>1}\dfrac{\gamma^2_{k0}}{\sqrt{2\xi\left(\xi-\gamma_{10}\right)}}
\\
&\quad
+\sum_{k>1}\delta_{jk}\gamma_{k0},
\end{split}
\ee
where, according to Eq.~(\ref{eq:xi_square}),
\be
\begin{split}
&\dfrac{(\gamma_{10}-\xi)\gamma_{10}}{\sqrt{2\xi\left(\xi-\gamma_{10}\right)}}
+\sum_{k>1}\dfrac{\gamma^2_{k0}}{\sqrt{2\xi\left(\xi-\gamma_{10}\right)}}
\\
&=\dfrac{1}{\sqrt{2\xi\left(\xi-\gamma_{10}\right)}}\left(\xi^2-\xi\gamma_{10}\right)
\\
&=\dfrac{1}{2}\sqrt{2\xi\left(\xi-\gamma_{10}\right)},
\end{split}
\ee
thus leading to 
\be
\begin{split}
P_{j1}\gamma_{10}+\sum_{k>1}P_{jk}\gamma_{k0}
&=\delta_{j1}\gamma_{10}-v_j\sqrt{2\xi\left(\xi-\gamma_{10}\right)}
\\
&\quad
+\sum_{k>1}\delta_{jk}\gamma_{k0},
\end{split}
\ee
and, consequently, to the final expression
\be\label{eq:gamma2_j0_general_exp}
\begin{split}
\left[\tilde{\bm\gamma}^2\right]_{j0}&\overset{j>0}{=}
n\left(\gamma_{10}\delta_{j1}-\sqrt{2\xi\left(\xi-\gamma_{10}\right)}v_j\right)
\\
&\quad+\sum_{k>1}\delta_{jk}\left(n\gamma_{k0}+\gamma_{10}\gamma_{k1}+\sum_{1<l\neq
k}\gamma_{kl}\gamma_{l0}\right)
\\
&\quad
+\left[P_{j1}-\dfrac{2v_j\gamma_{10}}{\sqrt{2\xi\left(\xi-\gamma_{10}\right)}}\right]
\sum_{k>1}\gamma_{k0}\gamma_{k1}
\\
&\quad
-\dfrac{4v_j}{\sqrt{2\xi\left(\xi-\gamma_{10}\right)}}\sum_{1<k<l}\gamma_{k0}\gamma_{kl}\gamma_{l0}.
\end{split}
\ee
As shown in the following, the last two terms on the right-hand side of
Eq.~(\ref{eq:gamma2_j0_general_exp}) vanish at half-filling ({\ie}, when
$n=1$). To prove this, let us consider the modified lattice Hamiltonian
\be\label{eq_appendix:part_hamilt}
\hat{H}\rightarrow
\hat{H}_{kl}(\eta)=\hat{H}+\eta\sum_\sigma\left(\hat{c}^\dagger_{k\sigma}\hat{c}_{l\sigma}
+\hat{c}_{l\sigma}^\dagger\hat{c}_{k\sigma}\right),
\ee   
where $0\leq k<l<L$ are {\it fixed} site labels. According to the
Hellmann--Feynman theorem, the exact $N$-electron density
matrix element for the Hubbard Hamiltonian $\hat{H}$ can be determined as follows:
\be\label{eq_appendix:1RDM_Hell-Feyn_th}
{\gamma}^N_{kl}
:=\left\langle\hat{c}^\dagger_{k\sigma}\hat{c}_{l\sigma}\right\rangle^N_{\hat{H}}=\dfrac{1}{4}\left.\dfrac{dE^N_{kl}(\eta)}{d\eta}\right|_{\eta=0},
\ee
where $E^N_{kl}(\eta)$ is the $N$-electron energy for the Hamiltonian
$\hat{H}_{kl}(\eta)$. If we now apply the following hole-particle
transformation~\cite{senjean2018site},
\be
\hat{c}_{i\sigma}\rightarrow
\hat{b}_{i\sigma}=(-1)^i\hat{c}_{i\sigma}^\dagger,
\ee 
the latter Hamiltonian can be rewritten as follows, 
\be\label{eq_appendix:hole_hamilt}
\begin{split}
\hat{H}_{kl}(\eta)&=
-t\sum_{\sigma}\sum^{L-1}_{ i= 0
}\left(\hat{b}^\dagger_{i\sigma}\hat{b}_{(i+1)\sigma}+
{\rm H.c.}
\right)
\\
&\quad
+U\sum^{L-1}_{i=0}\hat{b}^\dagger_{i\uparrow}\hat{b}_{i\uparrow}\hat{b}^\dagger_{i\downarrow}\hat{b}_{i\downarrow}
\\
&\quad+U\left(L-\sum_{\sigma}\sum^{L-1}_{ i= 0}\hat{b}^\dagger_{i\sigma}\hat{b}_{i\sigma}\right)
\\
&\quad+(-1)^{l-k-1}\eta\sum_\sigma\left(\hat{b}^\dagger_{k\sigma}\hat{b}_{l\sigma}
+\hat{b}_{l\sigma}^\dagger\hat{b}_{k\sigma}\right).
\end{split}
\ee
Note that
$\hat{b}_{L\sigma}=(-1)^L\hat{c}_{L\sigma}^\dagger=\pm(-1)^L\hat{c}^\dagger_{0\sigma}=\pm(-1)^L\hat{b}_{0\sigma}$,
which means that the same (periodic or antiperiodic) boundary conditions
can be used in the
particle [Eq.~(\ref{eq_appendix:part_hamilt})] or hole
[Eq.~(\ref{eq_appendix:hole_hamilt})] Hamiltonians for a total even
number $L$ of sites. In this case, we can conclude from
Eq.~(\ref{eq_appendix:hole_hamilt}) that   
\be
E^{2L-N}_{kl}(\eta)=E^N_{kl}\Big((-1)^{l-k-1}\eta\Big)+U(L-N),
\ee
thus leading to [see Eq.~(\ref{eq_appendix:1RDM_Hell-Feyn_th})]
\be
{\gamma}^{2L-N}_{kl}=
(-1)^{l-k-1}
{\gamma}^N_{kl}.
\ee
Therefore, in the half-filled $n=N/L=1$ case, we have
\be
{\gamma}^{N=L}_{kl}\times\left(1+(-1)^{l-k}\right)=0,
\ee
that we simply denote
\be
{\gamma}_{kl}\times\left(1+(-1)^{l-k}\right)\overset{n=1}{=}0.
\ee
In conclusion, at half-filling, a density matrix element equals
zero if the two indices differ by an even number:
\be\label{eq_appendix:1RDM_elt_zero_half-fill}
{\gamma}_{kl}&
\overset{n=1}{=}&0\;\;\;\mbox{if}\;\;\; l-k=2p>0.
\ee 
If we now return to Eq.~(\ref{eq:gamma2_j0_general_exp}), we immediately
see that, for $k>1$,
\be
\gamma_{k0}\gamma_{k1}\overset{n=1}{=}0
\ee 
and
\be
\gamma_{k0}\gamma_{kl}\gamma_{l0}\overset{n=1}{=}0,
\ee
so that
\be\label{eq:gamma2_j0_half-filling}
\begin{split}
&\left[\tilde{\bm\gamma}^2\right]_{j0}\overset{j>0,n=1}{=}
\left(\gamma_{10}\delta_{j1}-\sqrt{2\xi\left(\xi-\gamma_{10}\right)}v_j\right)
\\
&\quad+\sum_{k>1}\delta_{jk}\left(\gamma_{k0}+\gamma_{10}\gamma_{k1}+\sum_{1<l\neq
k}\gamma_{kl}\gamma_{l0}\right).
\end{split}
\ee
We can now extract interesting information from our simplified
Eq.~(\ref{eq:gamma2_j0_half-filling}). For example, for $j=1$, we
obtain, according to Eq.~(\ref{eq:Householder_vec}),
\be
\left[\tilde{\bm\gamma}^2\right]_{10}\overset{n=1}{=}\gamma_{10}-\sqrt{2\xi\left(\xi-\gamma_{10}\right)}v_1=\xi,
\ee
which, according to Eq.(\ref{eq:nbr_electrons_cluster}), leads to
$\mathcal{N}^{\mathcal{C}}=2$. Moreover, 
\be
\begin{split}
&\left[\tilde{\bm\gamma}^2\right]_{(2p+1)0}\overset{p>0,n=1}{=}
-\gamma_{(2p+1)0}+\gamma_{(2p+1)0}
\\
&\quad+\gamma_{10}\gamma_{(2p+1)1}+\sum_{1<l\neq
2p+1}\gamma_{(2p+1)l}\gamma_{l0},
\end{split}
\ee
thus leading to [see Eqs.~(\ref{eq:DM_buffer_region}) and (\ref{eq_appendix:1RDM_elt_zero_half-fill})]
\be
\tilde{\gamma}_{(2p+1)1}\overset{p>0,n=1}{=}0.
\ee
On the other hand, the Householder transformed density matrix elements
in column ``1"
are {\it a priori} nonzero for even row indices, 
\be
\begin{split}
\xi\tilde{\gamma}_{(2p)1}&\overset{p>0}=\left[\tilde{\bm\gamma}^2\right]_{(2p)0}
\overset{p>0,n=1}{=}\gamma_{10}\gamma_{(2p)1}+\sum_{1<l\neq
2p}\gamma_{(2p)l}\gamma_{l0}
\\
&\neq 0,
\end{split}
\ee
which means that the cluster is {\it not} disconnected from its environment, even
though it contains exactly two 
electrons.

\section{``Direct'' correlation effects on the Householder
transformation}\label{appendix:correlation_and_bath}

Let us consider the idempotent non-interacting density matrix $\bg$ in the lattice
representation, $\bP\equiv \bP[\bg]$ [see
Eqs.~(\ref{eq:Pij_exp})--(\ref{eq:xi_final_exp})] the corresponding
Householder transformation matrix, and
$\tilde{\bm\gamma}=\bP\bg\bP$. We want to see how the Householder vector
is {\it directly} affected (\ie, through modifications of the density
matrix in the lattice
representation) by changes 
\be
\tilde{\bm\gamma}\rightarrow \tilde{\bm\gamma}+{\bm \Delta}
\ee
that may occur in the Householder representation of the density matrix,
for example, when electron correlation is introduced. For that purpose,
we need to return to the lattice representation,
\be
\begin{split}
{\bm\gamma}\rightarrow\overline{\bm\gamma}&=\bP\left(\tilde{\bm\gamma}+{\bm
\Delta}\right)\bP
\\&=\bg+\bP{\bm\Delta}\bP
,
\end{split}
\ee
thus leading to
\be
\gamma_{i0}\rightarrow\overline{\gamma}_{i0}&=&\gamma_{i0}+\sum_jP_{ij}\Delta_{j0}
\\
\label{eq_appendix=details_X_withDelta}
&=&\gamma_{i0}+\Delta_{i0}-2v_i\sum_{j>0}v_j\Delta_{j0},
\ee
or, in a more compact form,
\be
\bX\rightarrow \overline{\bX}=\bX+\bP{\bm \Delta}_0,
\ee
where ${\bm
\Delta}_0^\dagger=\left[\Delta_{00},\Delta_{10},\ldots,\Delta_{j0},\ldots\right]$.
As readily seen from Eq.~(\ref{eq_appendix=details_X_withDelta}),
correlating within the Householder environment (which would imply
$\Delta_{j0}=0, \forall j$) has not impact on $\bX$ and,
therefore, no impact on the Householder transformation. However, 
changes occur on all rows of $\bX$ when the density matrix is
modified within the cluster, {\ie}, when
$\Delta_{j0}=\sum^1_{i=0}\delta_{ij}\Delta_{i0}$. As we will see, even
though $\bX$ and $\bY=\bP\bX$ change in this case, $\bX-\bY$ will not change
direction.\\

To prove the above statement, we need to evaluate the change
$\bY^\dagger\rightarrow\overline{\bY}^\dagger=\left[\overline{\gamma}_{00},\overline{\xi},0,\ldots,0\right]$ in $\bY$.
Since, by construction,    
\be
\abs{\overline{\bY}}^2=\abs{\overline{\bX}}^2=\abs{\bP\overline{\bX}}^2=\abs{\bY+{\bm
\Delta}_0}^2,
\ee
it comes from Eq.~(\ref{eq_appendix=details_X_withDelta}) that
\be
\begin{split}
\overline{\xi}^2
&=\abs{\overline{\bY}}^2-\overline{\gamma}^2_{00}
\\&=\sum_{i}\left(Y_i+\Delta_{i0}\right)^2-\left(\gamma_{00}+\Delta_{00}\right)^2
\\
&=\left(\xi+\Delta_{10}\right)^2+\sum_{i>1}\Delta^2_{i0},
\end{split}
\ee
or, equivalently,
\be
\overline{\xi}=\left(\xi+\Delta_{10}\right)\sqrt{1+\dfrac{\sum_{i>1}\Delta^2_{i0}}{\left(\xi+\Delta_{10}\right)^2}},
\ee
thus leading to
\be\label{eq_appendix:xi_diff_final_exp}
\begin{split}
\overline{\xi}-\xi&=\xi\left(\sqrt{1+\dfrac{\sum_{i>1}\Delta^2_{i0}}{\left(\xi+\Delta_{10}\right)^2}}-1\right)
\\
&\quad+\Delta_{10}\sqrt{1+\dfrac{\sum_{i>1}\Delta^2_{i0}}{\left(\xi+\Delta_{10}\right)^2}}.
\end{split}
\ee
Note that the sign of $\overline{\xi}$ has been chosen such that
$\overline{\xi}=\xi$ when ${\bm \Delta}_0=0$, as expected. 
Since
\be
\begin{split}
\overline{\bX}-\overline{\bY}&=\overline{\bX}-\bY-\left(\overline{\bY}-\bY\right)
\\
&=\left(\bX-\bY\right)+\bP{\bm \Delta}_0-\Delta_{00}\bev_0
\\
&\quad
+\left(\xi-\overline{\xi}\right)\bev_1,
\end{split}
\ee
where $\bev_i$ are the unit basis vectors ($\bev_0^\dagger=[1,0,0,\ldots,0]$,
$\bev_1^\dagger=[0,1,0,\ldots,0]$),
we can now determine the direction of the updated Householder vector:
\be\label{eq_appendix:new_HH_vector}
\begin{split}
\overline{\bv}\sim \overline{\bX}-\overline{\bY}&=\left(\abs{\bX-\bY}-2\bv^\dagger{\bm
\Delta}_0\right)\bv
\\
&\quad
+\sum_{i>0}\left[\Delta_{i0}+\delta_{i1}\left(\xi-\overline{\xi}\right)\right]\bev_i.
\end{split}
\ee
As readily seen from Eqs.~(\ref{eq_appendix:xi_diff_final_exp}) and (\ref{eq_appendix:new_HH_vector}), 
when correlation is treated within the Householder cluster only (\ie,
when
$\Delta_{i0}\overset{i>1}{=}0$), then 
$\overline{\xi}-\xi=\Delta_{10}$ and
\be
\overline{\bv}\sim \left(\abs{\bX-\bY}-2v_1{
\Delta}_{10}\right)\bv\sim \bv,
\ee
which, after normalization, leads to $\overline{\bv}=\bv$ and
$\overline{\bP}=\bP$.

\end{appendices}




\newcommand{\Aa}[0]{Aa}
\end{document}